\RequirePackage{amsthm}
\documentclass[default,iicol,sn-mathphys,Numbered]{sn-jnl}
\usepackage[utf8]{inputenc}
\usepackage[T1]{fontenc}
\usepackage{makecell}
\usepackage{cellspace}
\usepackage{colortbl}
\usepackage{graphicx}
\usepackage{multirow}
\usepackage{amsmath,amssymb,amsfonts}
\usepackage[title]{appendix}%
\usepackage{textcomp}%
\usepackage{url}
\usepackage{xcolor}
\usepackage{manyfoot}
\usepackage{booktabs}
\usepackage{algorithm}
\usepackage{algorithmicx}
\usepackage{algpseudocode}
\usepackage{listings}
\usepackage{hyperref}
\usepackage{geometry}
\usepackage{svg}
\usepackage{lmodern}
\usepackage[normalem]{ulem}
\useunder{\uline}{\ul}{}
\usepackage[switch]{lineno}
\begin{document}
%\linenumbers

\title[Article Title]{Free Lunch in Medical Image Foundation Model Pre-training via Randomized Synthesis and Disentanglement}

\author[1]{\fnm{Yuhan} \sur{Wei}}\email{yweibs@connect.ust.hk}
\equalcont{These authors contributed equally to this work.}

\author[2]{\fnm{Yuting} \sur{He}}\email{yuting.he4@case.edu}
\equalcont{These authors contributed equally to this work.}

\author[1]{\fnm{Linshan} \sur{Wu}}\email{linshan.wu@connect.ust.hk}

\author[3]{\fnm{Fuxiang} \sur{Huang}}\email{fxhuang1995@gmail.com}

\author[1]{\fnm{Junlin} \sur{Hou}}\email{csejlhou@ust.hk}

\author*[1,4,6,7]{\fnm{Hao} \sur{Chen}}\email{jhc@ust.hk}

\affil[1]{Department of Computer Science and Engineering, the Hong Kong University of Science and Technology, Hong Kong, China}
\affil[2]{Department of Biomedical Engineering, Case Western Reserve University, OH, USA}
\affil[3]{School of Data Science, Lingnan University, Hong Kong, China}
\affil[4]{Department of Chemical and Biological Engineering and Division of Life Science, Hong Kong University of Science and Technology, Hong Kong, China}
\affil[6]{HKUST Shenzhen-Hong Kong Collaborative Innovation Research Institute, Futian, Shenzhen, China}
\affil[7]{State Key Laboratory of Nervous System Disorders, The Hong Kong University of Science and Technology, Hong Kong, China}

\abstract{Medical image foundation models (MIFMs) have demonstrated remarkable potential for a wide range of clinical tasks, yet their development is constrained by the scarcity, heterogeneity, and high cost of large-scale annotated datasets. Here, we propose RaSD (Randomized Synthesis and Disentanglement), a scalable framework for pre-training MIFMs entirely on synthetic data. By modeling anatomical structures and appearance variations with randomized Gaussian distributions, RaSD exposes models to sufficient multi-scale structural and appearance perturbations, forcing them to rely on invariant and task-relevant anatomical cues rather than dataset-specific textures, thereby enabling robust and transferable representation learning. We pre-trained RaSD on 1.2 million 3D volumes and 9.6 million 2D images, and extensively evaluated the resulting models across 6 imaging modalities, 48 datasets, and 56 downstream tasks. Across all evaluated downstream tasks, RaSD consistently outperforms training-from-scratch models, achieves the best performance on 17 tasks, and remains comparable to models pre-trained on large real datasets in most others. These results demonstrate that the capacity of synthetic data alone to drive robust representation learning. Our findings establish a paradigm shift in medical AI, demonstrating that synthetic data can serve as a ``free lunch'' for scalable, privacy-preserving, and clinically generalizable foundation models.
}

\maketitle

\section{Introduction}\label{sec:intro}
Medical image foundation models (MIFMs), pre-trained on large and diverse datasets, have emerged as a promising paradigm for a broad spectrum of medical image tasks, with increasing relevance to clinical translation \cite{bommasani2021opportunities,he2024foundation,ma2025fully,sun2025data,chen2024towards}. Through large-scale pre-training (Fig.~\ref{fig:intro}-b), MIFMs acquire transferable visual representations that can be efficiently adapted to specific downstream applications \cite{bommasani2021opportunities}, conferring two major advantages: 1) Enhanced performance: exposure to diverse data distributions during pre-training enables MIFMs to encode domain-invariant features, improving accuracy and robustness in fine-tuned tasks. 2) Reduced training costs: their generalization capacity reduces the need for extensive task-specific annotations and computational overhead, thereby streamlining model development. Such advantages have positioned MIFMs as a versatile tool for critical medical image applications \cite{chen2024towards,ma2025fully,sun2025data} with growing potential for clinical integration.

\begin{figure*}
    \centering
    \includegraphics[width=1\linewidth]{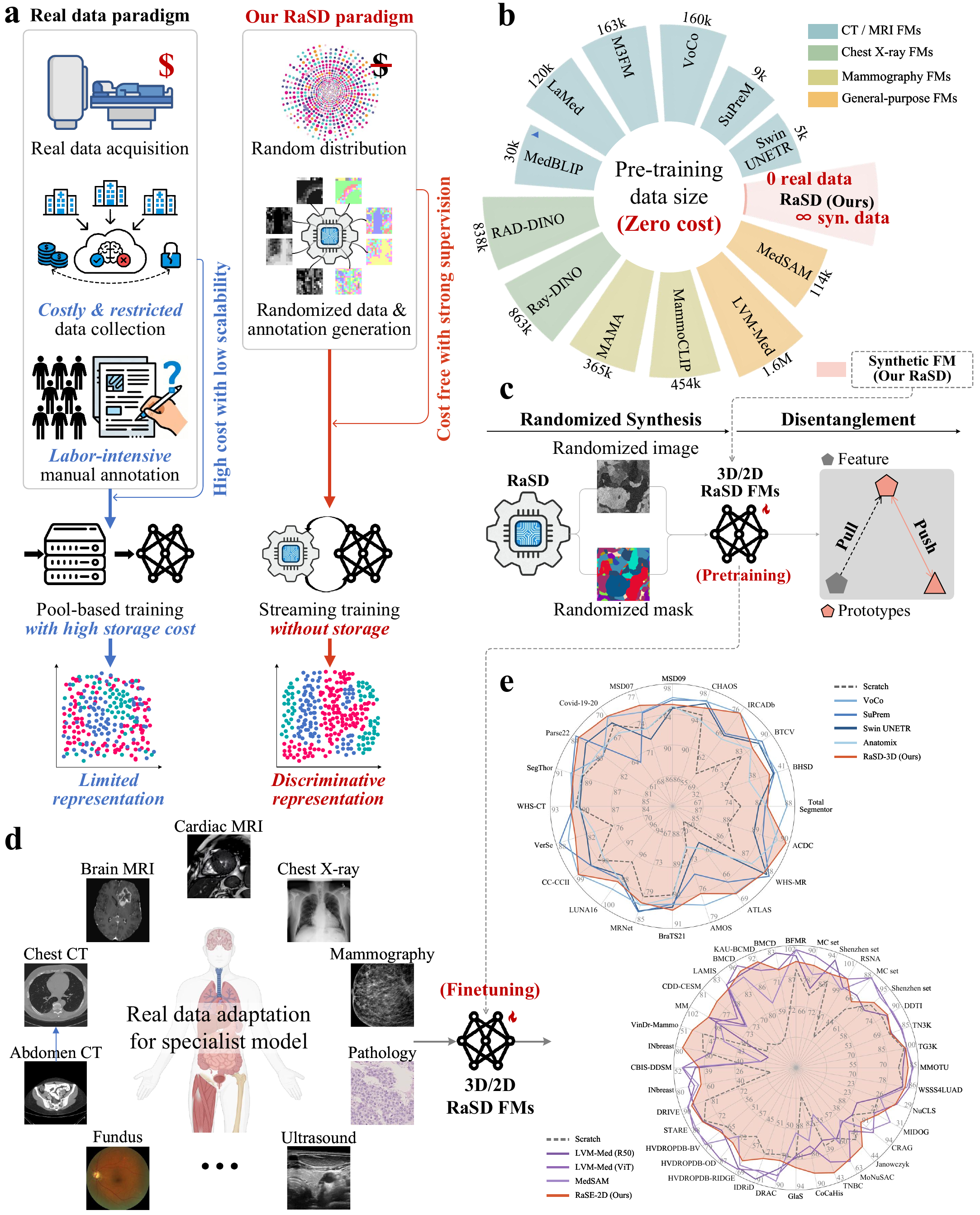}
\end{figure*}
\begin{figure*}
    \centering
    \caption{Overview of the proposed RaSD framework. a) Comparison between the conventional real data paradigm and our RaSD paradigm. Real data-based pre-training of MIFMs suffers from costly and restricted data acquisition, labor-intensive manual annotation, and large storage demands, leading to limited discriminative capacity. In contrast, RaSD synthesizes diverse image–label pairs from randomized distributions in a streaming manner, requiring no pre-generated data storage. b) Pre-training data scale of existing foundation models compared with RaSD. Unlike prior FMs that depend on large-scale real-image datasets, RaSD achieves large-scale pre-training solely from synthetic data at zero real-data cost. c) Core components of RaSD. Randomized images and masks are generated on-the-fly, and the model is pre-trained to disentangle structural features and learn discriminative representations. d) Real-data adaptation. RaSD-pretrained FMs can be efficiently fine-tuned across diverse modalities (e.g., MRI, CT, X-ray, mammography, pathology and fundus), enabling efficient adaptation to task-specific models. e) Benchmarking results. RaSD-pretrained models (2D/3D) achieve competitive or superior performance compared with state-of-the-art foundation models across 48 datasets and 56 downstream tasks, demonstrating strong transferability and robustness.}
    \label{fig:intro}    
\end{figure*}

Despite promising advances in MIFMs \cite{tiu2022expert,zhou2023foundation,ma2025fully,yan2025multimodal}, data remains a central bottleneck in their pre-training (Fig.~\ref{fig:intro}-a). The acquisition of medical images is both expensive and constrained by regulatory and technical barriers \cite{kayaalp2018patient}. Variability across institutions, imaging devices, and anatomical sites further exacerbates data heterogeneity \cite{rahman2013addressing}. These factors extremely increase the cost of medical data collection, limiting the scalability and diversity of datasets required for model generalization across diverse clinical scenarios \cite{bommasani2021opportunities}. In addition, high-resolution and volumetric data from modalities like Computed Tomography (CT), Magnetic Resonance Imaging (MRI), and whole slide imaging (WSI) require significant storage space and high bandwidth for retrieval. Supervised learning \cite{li2024well} becomes impractical due to the high cost of manual annotation for large-scale datasets, while self-supervised learning \cite{zhou2021models,he2023geometric,wu2024voco} struggles to provide sufficiently strong supervision signals, failing to disentangle complex semantic information in medical images. Collectively, these limitations impair the ability of MIFMs to learn transferable and discriminative features across diverse clinical contexts.

Recent advances have underscored the growing potential of synthetic data in model training \cite{fan2024scaling,sheng2025synthetic,sun2025data,wang2025self}. Synthetic data can be generated efficiently and flexibly, producing diverse images with rich annotations at minimal cost. By eliminating concerns over data privacy \cite{kayaalp2018patient}, manual annotation, and storage costs, synthetic data presents a promising solution for the challenges of MIFM pre-training. Synthetic images can be generated on demand with automatically assigned labels, enabling scalable pre-training without relying on costly real datasets or risking patient privacy. A growing body of work has focused on training generative foundation models (FMs) to produce synthetic data tailored to specific downstream tasks \cite{sheng2025synthetic,sun2025data,wang2025self}. However, these generative FMs still depend heavily on large quantities of real data during training, which may be costly, sensitive, or limited in coverage, and thus cannot fully escape the practical constraints of real-world data. In contrast, as shown in Fig.\ref{fig:intro}-a, rule-based or randomized synthesis approaches \cite{billot2023synthseg,hoffmann2021synthmorph,dey2025learning} suggest the possibility of constructing synthetic datasets without real data input, particularly in medical imaging domains where structure and anatomy follow relatively well-defined patterns. These observations lead us to formulate a hypothesis: ``\textit{Synthetic medical image data may serve as a free lunch for MIFM pre-training, making it possible to train MIFMs without real data.}'' Although a preliminary work\cite{dey2025learning} has explored randomized synthesis for MIFM pre-training, their method focuses on enforcing appearance consistency, limiting the tasks where appearance is an important feature.

In this study, we propose a scalable synthetic data-driven pre-training framework for MIFM, \textbf{Ra}ndomized \textbf{S}ynthetic data \textbf{D}isentanglement (\textbf{RaSD}, Fig.\ref{fig:intro}-c,d), which operates entirely without reliance on real medical data. RaSD synthesizes diverse image–label pairs by sampling from randomized statistical distributions that capture common structural and appearance variations in medical images, eliminating the requirement for physical imaging devices and mitigating privacy risks. Its online synthesis pipeline enables streaming training, avoiding the storage of large datasets. Each synthetic image is generated alongside pixel-wise annotations, obviating manual labeling and facilitating efficient supervised training. By exposing models to large-scale and diverse synthetic data, RaSD fosters the learning of transferable features that generalize beyond the limitations of real data scarcity. This framework establishes a new paradigm in medical AI, liberating MIFM pre-training from real data, and paving the way for generalizable and scalable models across numerous clinical scenarios.

Our RaSD framework was pre-trained on an unprecedented scale of 1.2 million (M) synthetic 3D volumes and 9.6M synthetic 2D images, enabling MIFMs of varying capacities to be benchmarked across both 2D and 3D settings, 6 imaging modalities, 48 datasets, and 56 downstream tasks (Fig.\ref{fig:intro}-e). Across this extensive evaluation, RaSD-pretrained models consistently matched or surpassed those trained on real medical datasets or natural images with striking gains. This effectiveness stems from RaSD's unique properties. It enables zero-cost data generation without reliance on imaging devices or privacy-sensitive collections, and supports storage-free online synthesis, where data are generated and consumed in real time. Each synthetic image is accompanied by automatic pixel-wise annotation, eliminating the burden of manual labeling, while large-scale and diverse synthetic distributions foster robust and discriminative representation learning. Collectively, these results provide compelling evidence that synthetic data alone can sustain competitive representation learning, establishing RaSD as a transformative foundation for scalable, privacy-preserving, and generalizable medical image analysis.

\section{Results}
\begin{figure*}
    \centering
    \includegraphics[width=0.94\textwidth]{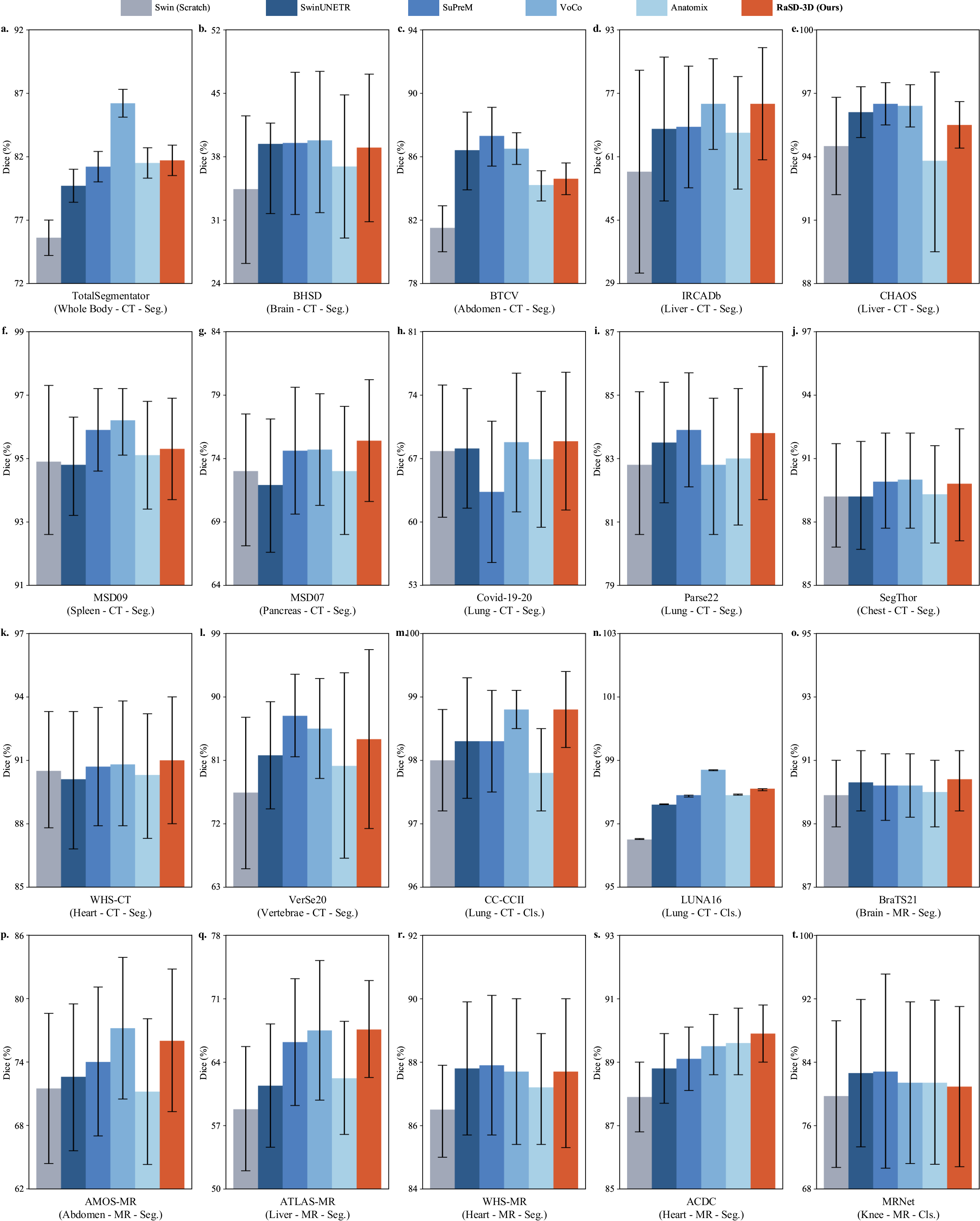}
    \caption{Our evaluations across diverse 3D radiology downstream tasks (CT and MR) demonstrate the strong transferability of RaSD. For CT tasks (\textbf{a–n}), RaSD covers multiple organs and anatomical structures across different body regions (e.g., abdomen, chest, and vertebrae), delivering competitive performance despite being trained solely on synthetic data, highlighting its potential as a universal MIFM pre-training strategy. For MR tasks (\textbf{o–t}), spanning multiple organs and anatomical regions (e.g., brain, heart and knee), RaSD likewise achieves stable gains. RaSD achieves consistent effectiveness across both CT and MR modalities, validating its ability to capture transferable cross-modal knowledge from synthetic data.}
    \label{fig:Comparison1}
\end{figure*}

\subsection{RaSD: MIFM pretraining with large-scale synthetic data} 
Our Randomized Synthetic data Disentanglement (RaSD) enables the pre-training of MIFMs entirely on large-scale synthetic data, without reliance on real images or manual annotation. During training, RaSD dynamically generates randomized anatomical labels and their corresponding images, providing paired supervision while eliminating the labor-intensive annotation process and the storage burden of conventional datasets. Owing to the negligible cost of machine-generated data, RaSD can continuously produce an effectively unlimited stream of synthetic samples for pre-training. In this work, we trained large 3D and 2D MIFMs using a total of 1.2M synthetic 3D image–label pairs and 9.6M synthetic 2D image–label pairs, establishing a scalable and storage-free paradigm for FM development in medical images.

The effectiveness of RaSD arises from our designed synthesis rules for medical imaging and the learning strategy for disentangling semantic structures. Grounded in the observation that medical images consist of multiple anatomical structures that can be effectively modeled using Gaussian distributions \cite{mclachlan2000finite,zhang2001segmentation,reynolds2009gaussian}, RaSD generates synthetic images by sampling Gaussian-distributed seeds and simulating imaging patterns from Gaussian noise with varied parameters, training models to disentangle intrinsic regions so that they acquire the fundamental ability to extract semantic structures and transfer this knowledge to downstream tasks.

To evaluate the scalability of RaSD under different dimensional settings, we pre-trained two MIFMs in both 2D and 3D: a 2D model based on a CNN encoder implemented in a 2D UNet \cite{ronneberger2015u} architecture (RaSD-2D), and a 3D model based on a Swin Transformer (base) \cite{liu2021swin} implemented within a Swin UNETR \cite{tang2022self} architecture (RaSD-3D). During training, RaSD dynamically generated 9.6M synthetic 2D image–label pairs for pre-training the 2D model, and 1.2M randomized 3D image–label pairs for the streaming pre-training of the 3D model, surpassing the size of most real-world datasets (Fig.~\ref{fig:intro}-c). This large-scale pre-training highlights RaSD's ability to efficiently scale across model sizes and architectures, establishing a robust foundation for assessing downstream performance across diverse medical imaging tasks. More details about our method and experiments are described in the ``Methods'' section~\ref{sec:method}.

\subsection{RaSD improves model generalization in 3D CT and MRI images}
Our RaSD-3D exhibits strong performance across a wide spectrum of CT and MR downstream tasks, demonstrating that synthetic data pre-training can effectively optimize CT and MR representations. We benchmarked our RaSD-3D on 14 CT datasets and 6 MR datasets spanning nearly all human organs, covering 17 segmentation and 3 classification tasks. Comparisons were made against three state-of-the-art (SOTA) self-supervised pre-training models (Swin UNETR (pretrained) \cite{tang2022self}, SuPreM \cite{li2024well}, VoCo \cite{wu2024voco,wu2024large}), a from-scratch baseline (Swin transformer (scratch) \cite{hatamizadeh2021swin}), and a most-related randomized synthesis-based FM (Anatomix \cite{dey2025learning}). For segmentation, their backbones were integrated into Swin UNETR, while for classification, the backbones were paired with a classification head following \cite{wu2024large}. Dice score and AUC were used for the evaluation of segmentation and classification tasks.

\subsubsection{CT image segmentation}
On CT datasets (Fig.~\ref{fig:Comparison1} a-l), our RaSD demonstrates strong generalization capabilities across anatomical regions. Compared with random initialization (``scratch''), our RaSD yields consistent and substantial improvements on all 12 CT segmentation tasks. Especially, on the TotalSegmentator (a), IRCADb (d), and VerSe20 (l), it brings over 6\% improvement, showing the effectiveness of our learning on synthetic images. Against real-data pre-training methods, RaSD achieves the best or tied-best performance on IRCADb (d), Covid-19-20 (h), and WHS-CT (k). Consistent performance is observed across diverse CT segmentation tasks, including BHSD (b), CHAOS (e), MSD09 (f), Parse22 (i), and SegThor (j), with RaSD performing within 1\% of the best method. 

Compared with Anatomix, our RaSD outperformed it in all 12 CT segmentation tasks, demonstrating the effectiveness of learning through randomized synthesis and disentanglement. While Anatomix also employs randomized data for FM training, its design primarily benefits modalities with unstable intensity distributions, which constrains its performance on intensity-stable modalities. Consequently, Anatomix shows markedly limited performance on the TotalSegmentator and BHSD datasets, in some cases performing worse than models with random initialization. In contrast, RaSD consistently achieves high performance across all datasets, highlighting its robustness and generalizability across diverse CT modalities.

\subsubsection{MR image segmentation}
On MR datasets (Fig.~\ref{fig:Comparison1} o–s), our RaSD continues to yield strong gains without requiring real-data pre-training. It attains the highest performance on 3 out of 5 tasks (o, q, s), outperforming all real-data pre-trained FMs. Moreover, it consistently outperforms at least one comparative method on the remaining two tasks (p, q). These results collectively indicate that the representations acquired through synthetic data pre-training not only transfer effectively to MR modalities but also exhibit particular advantages in segmenting diverse anatomical structures, including brain tumors, whole-body organs, and cardiac anatomy.

Compared with Anatomix, our RaSD also achieved superior performance in MR segmentation. Although Anatomix demonstrates comparable results in brain, liver, and heart segmentation tasks (o, q, r, s), its performance on abdominal multi-organ segmentation (AMOS-MR, p) is severely limited, in some cases falling below that of models with random initialization. This limitation arises from Anatomix’s suppression of appearance differences, which restricts the representation of organ and structural features and reduces feature discriminability. In contrast, RaSD consistently delivers stable performance gains, even outperforming Swin UNTER, which is pre-trained on real images.

\subsubsection{CT/MR classification}
Beyond segmentation, RaSD also demonstrates compelling cross-task generalization in classification datasets (Fig.~\ref{fig:Comparison1} m, n, t). 

On CT-based datasets, RaSD achieves an AUC of 98.8\% on CC-CCII (m), matching VoCo and surpassing SuPreM and Swin-UNETR by up to 0.5\%. On LUNA16 (n), RaSD attains 98.1\%, outperforming all other methods except VoCo, with only a small gap of 0.6\% compared to the best method. For the MR image benchmark (t), RaSD reaches 80.9\% in AUC, surpassing from scratch baseline by 1.2\% and closing the performance gap with top-performing methods within 2\%. These results confirm that RaSD not only overcomes the appearance constraints of Anatomix but also establishes a more robust and generalizable synthesis-based pre-training paradigm through randomized disentanglement.

\subsection{RaSD effectively generalizes in 2D X-ray image analysis}
\begin{figure*}
    \centering
    \includegraphics[width=0.94\textwidth]{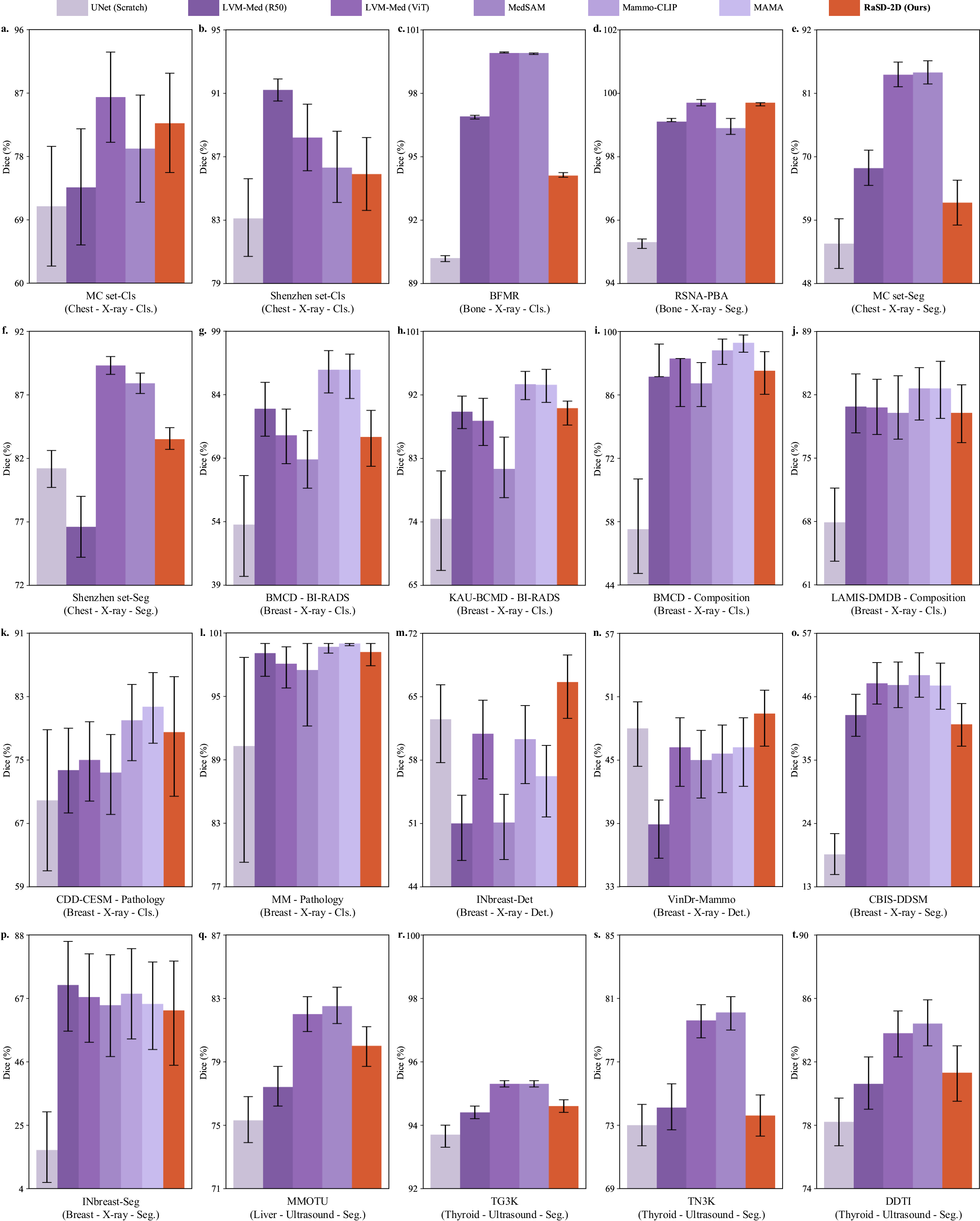}
    \caption{Our evaluations across 16 X-ray datasets and 4 ultrasoud datasets demonstrate the broad applicability of RaSD. For chest and skeletal X-rays (a–f), RaSD achieves competitive results on classification (a–c) and segmentation (d–f), confirming its strong transferability across major diagnostic applications. For mammography (g–p), spanning classification (g–l), detection (m, n), and segmentation (o, p), RaSD consistently delivers strong generalization despite being trained solely on synthetic data. For ultrasound tasks (q-t), RaSD surpasses from scratch model and is comparative to other methods. These results highlight the robustness of RaSD in capturing transferable representations across diverse X-ray sub-modalities and task types. The ``Composition'', ``BI-RAID'', and ``Pathology'' are the three standard criteria in breast cancer diagnosis \cite{fowler2013breast}. }
    \label{fig:Comparison2}
\end{figure*}

Our RaSD-2D also exhibits strong performance across a broad spectrum of X-ray downstream tasks, showing that synthetic data pre-training can effectively optimize X-ray representations. We benchmarked RaSD on 16 tasks spanning mammography, chest X-ray, and skeletal X-ray, covering classification, segmentation, and detection. Comparisons were made against SOTA general-purpose FMs (LVM-Med (R50/ViT) \cite{mh2023lvm}), supervised pre-trained FMs (MedSAM \cite{ma2024segment}), and modality-specific FMs (Mammo-CLIP \cite{ghosh2024mammo}, MAMA \cite{du2025multi}) to comprehensively compare and analyze the effectiveness of our RaSD-2D. Because the Anatomix \cite{dey2025learning} lacks a 2D solution, we did not include it in the validation of 2D scenes. For segmentation and detection, Dice and IoU were adopted as metrics, while AUC was used for classification.

\subsubsection{Mammography image analysis}
We evaluated the effectiveness of RaSD on mammography images (Fig.~\ref{fig:Comparison2} g–p), comparing it with general-purpose FMs, supervised pre-trained FMs, and modality-specific FMs. Across both segmentation and object detection tasks (m–p), RaSD achieved substantial performance gains. In detection tasks (m, n), it outperformed all competitors, including Mammo-CLIP and MAMA, by 4.1\% on INbreast and 1.4\% on VinDr-Mammo. In segmentation tasks (o, p), RaSD showed marked improvements over models trained from scratch and achieved performance comparable to other FMs pre-trained on real images. These gains can be attributed to our disentanglement learning, which naturally enhances dense feature discrimination, thereby facilitating downstream dense prediction tasks, i.e., segmentation and detection.

In breast cancer diagnosis tasks (g–k), RaSD demonstrated consistently strong performance. It outperformed general-purpose FMs in four out of six tasks and even surpassed the modality-specific FM on LAMIS-DMDB. In pathology diagnosis (k, l), RaSD exceeded general-purpose models by 3.5–5.0\%, while trailing specialized mammography models by only 1.6–3.3\%. In mammographic breast composition and BI-RADS classification (g–j), it remained competitive, consistently narrowing the gap with modality-specific baselines. Although RaSD was slightly behind Mammo-CLIP and MAMA in most experiments, the performance gap was consistently small—below 5\% in 80\% of comparisons and within 3\% in 60\%. This narrow margin, also observed in other general-purpose models, underscores RaSD's strong generalization: despite never encountering mammography data during pre-training, it effectively transfers universal knowledge to this modality.

\subsubsection{Radiography image analysis}
On chest and skeletal radiography tasks (Fig.~\ref{fig:Comparison2} a–f), RaSD consistently outperforms models trained from scratch, demonstrating the effectiveness of synthetic data pre-training. For chest radiography classification (a–b), RaSD improves the from-scratch baseline by 2.8\% on the Shenzhen set and 11.8\% on the MC set. Notably, on the MC set, RaSD even surpasses both LVM-Med (R50) and MedSAM, highlighting its strong representation capacity. For skeletal radiography fracture classification (c), RaSD attains an AUC of 94.1\%, exceeding scratch by 3.9\%. In chest and skeletal radiography segmentation (d–f), RaSD achieves robust improvements over the scratch model across all datasets and records best performance on RSNA-PBA (d). Taken together, these results demonstrate that synthetic data pre-training enables RaSD to consistently surpass from-scratch models, while remaining competitive with foundation models pre-trained on large amount of real medical images.

\subsection{RaSD generalizes effectively to ultrasound segmentation}
On ultrasound segmentation tasks (Fig.~\ref{fig:Comparison2} q-t), RaSD demonstrates strong and consistent generalization across diverse datasets, despite being pre-trained solely on synthetic data. Compared with the from-scratch baseline, RaSD achieves clear and statistically significant improvements on all four datasets, with improvements ranging from 0.6\% to 4.7\%. In particular, RaSD yields substantial improvements on the DDTI (t) and MMOTU (q) datasets, outperforming the scratch model by 3.1\% and 4.7\% respectively, indicating its effectiveness in learning robust structural representations under complex appearance variations and low-contrast conditions. Compared with real-data pre-trained models, RaSD also exhibits competitive performance. It surpasses LVM-Med (R50) on three out of four datasets (q, r, t), demonstrating that synthetic data pre-training can match and even exceed conventional real-data-based approaches in multiple scenarios. Although RaSD slightly underperforms LVM-Med (R50) on TN3K (s), the overall results indicate its strong and stable generalization capability. Furthermore, RaSD maintains a relatively small performance gap with the best-performing methods on most datasets. On TG3K (r), the gap to the top models is only 0.7\%. These results suggest that RaSD can closely match models pre-trained on large-scale real datasets, even without access to real images during pre-training.

\subsection{RaSD transfers robustly to 2D retinal image analysis}
\begin{figure*}
    \centering
    \includegraphics[width=0.94\textwidth]{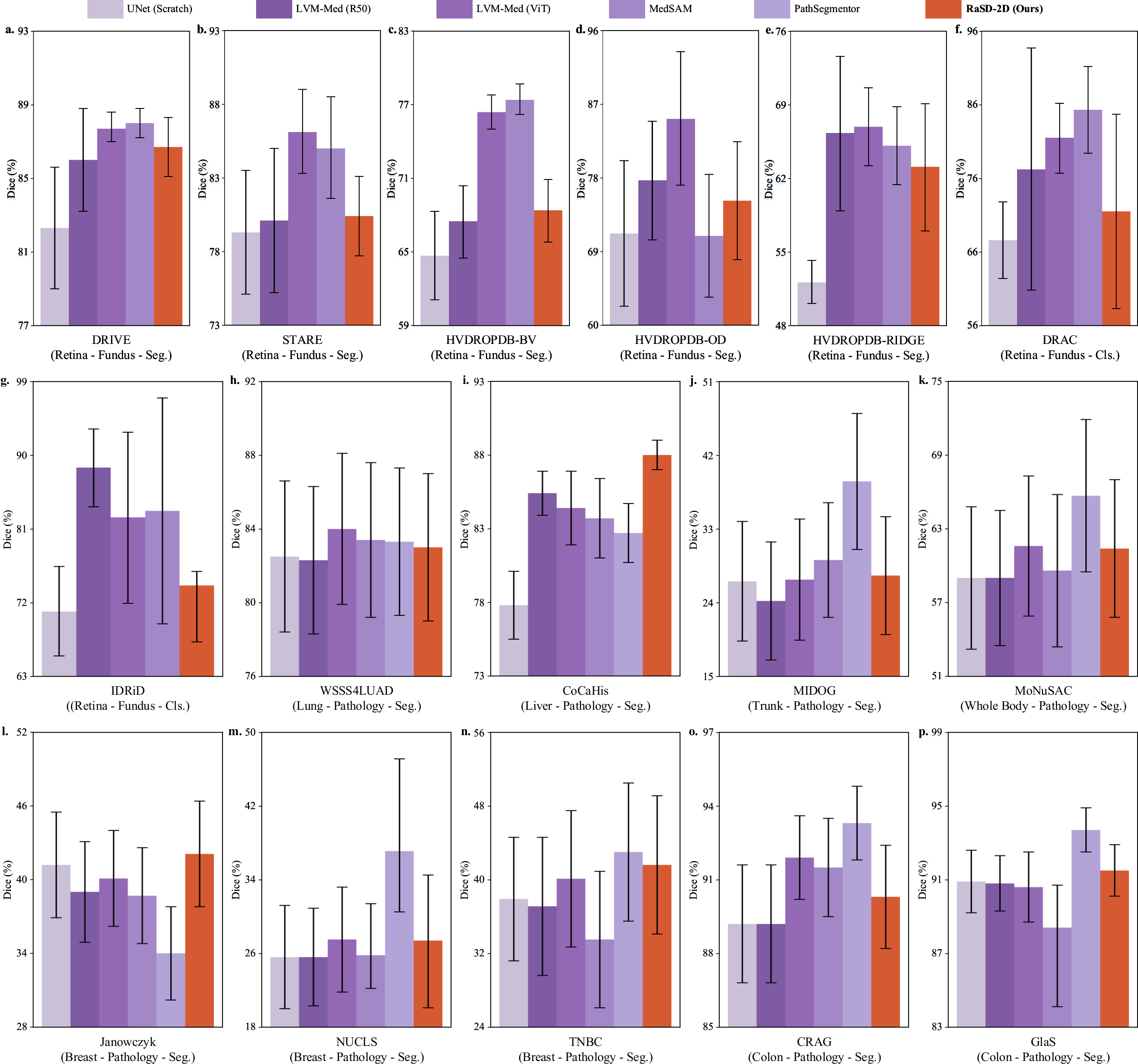}
    \caption{ RaSD demonstrates robust generalization to both 2D fundus and pathology tasks. Evaluations span Fundus (\textbf{a–g}) and pathology (\textbf{h-p}). For fundus segmentation tasks (\textbf{a–e}), spanning vessel, optic disc, and ridge structures across five benchmark datasets, RaSD achieves competitive Dice performance compared with SOTA FMs. For fundus classification tasks (\textbf{f-g}), covering retinal disease identification, RaSD likewise delivers stable AUC improvements over the scratch baseline and matches or surpasses real-data pre-trained models. Across nine pathology segmentation datasets (\textbf{h-p}), RaSD consistently matches or surpasses the performance of specialized models, achieving SOTA results on key datasets such as CoCaHis and Janowczyk. Collectively, these results across fundus and pathology modality validate RaSD's capacity to learn broadly transferable visual representations from synthetic data.}
    \label{fig:Comparison3}
\end{figure*}

\begin{table*}[htbp]
\centering
\caption{Comparative Analysis of Offline and Online Data Synthesis Paradigms Across Different Data Scales}
\label{tab:online_comparison}
\begin{tabular}{l l c c c c}
\toprule
\multirow{2}{*}{\textbf{Data Amount}} & 
\multirow{2}{*}{\textbf{\parbox{3cm}{\centering Data Synthesis Paradigm}}} & 
\textbf{\parbox{2.5cm}{\centering Data Preparation Time}} & 
\textbf{\parbox{2.5cm}{\centering Model Training Time}} & 
\textbf{Total Time} & 
\textbf{Storage} \\
\addlinespace
& & \textbf{(s)} & \textbf{(s)} & \textbf{(s)} & \textbf{(GB)} \\
\midrule

100 & Offline & 7.07 & 84.76 & 91.83 & 1.00 \\
    & Online  & 2.48 & 82.38 & 84.86 & 0.00 \\
\rowcolor{gray!15}
    & $\Delta$(Online-Offline) & $\downarrow$4.59 & $\downarrow$2.38 & $\downarrow$6.97 & $\downarrow$1.00 \\

200 & Offline & 17.88 & 147.19 & 165.07 & 1.96 \\
    & Online  & 4.54  & 138.60 & 143.14 & 0.00 \\
\rowcolor{gray!15}
    & $\Delta$(Online-Offline) & $\downarrow$13.34 & $\downarrow$8.59 & $\downarrow$21.93 & $\downarrow$1.96 \\

500 & Offline & 42.64 & 345.57 & 388.21 & 4.87 \\
    & Online  & 9.95  & 327.49 & 337.44 & 0.00 \\
\rowcolor{gray!15}
    & $\Delta$(Online-Offline) & $\downarrow$32.69 & $\downarrow$18.08 & $\downarrow$50.77 & $\downarrow$4.87 \\
    
1000 & Offline & 75.51 & 667.67 & 743.18 & 9.76 \\
     & Online  & 19.86 & 627.05 & 646.91 & 0.00 \\
\rowcolor{gray!15}
     & $\Delta$(Online-Offline) & $\downarrow$55.65 & $\downarrow$40.62 & $\downarrow$96.27 & $\downarrow$9.76 \\

\bottomrule
\end{tabular}
\end{table*}

Our RaSD also brings strong and consistent performance improvement across multiple fundus image tasks. We conducted experiments on five segmentation and three classification tasks and benchmarked RaSD against three SOTA MIFMs, including LVM-Med (R50), LVM-Med (ViT), and MedSAM, and a randomly initialized model (``Scratch'').
\subsubsection{Fundus image Segmentation}
RaSD demonstrates strong capability in retinal tissue segmentation across multiple datasets (Fig.~\ref{fig:Comparison3} a–e). 
On the STARE dataset (b), it achieved an 80.4\% Dice, surpassing LVM-Med (R50) by 0.3\%, underscoring its ability to capture fine-grained, discriminative features essential for intricate vessel structures. On vessel segmentation (DRIVE, a), RaSD attained an 86.7\% Dice, remaining within 1.5\% of the best-performing methods. Its robust performance further extends to optic disc and ridge segmentation (HVRODPDB-BV, HVRODPDB-OD, HVRODPDB-RIDGE; c–e), where it consistently outperformed scratch models, confirming its generalizability across diverse segmentation challenges in fundus imaging. These results highlight RaSD's ability to learn fine-grained and dense structural representations from synthetic data, enabling competitive performance in retinal segmentation tasks without reliance on real-image pre-training.

\subsubsection{Fundus image Classification}
RaSD also demonstrates competitive performance on retinal disease classification tasks (Fig.~\ref{fig:Comparison3} f-g). Across both DRAC and IDRiD datasets, RaSD consistently outperforms the from-scratch baseline, achieving statistically significant improvements of 3.9\% and 3.2\%, respectively, indicating that synthetic pre-training provides effective initialization for fundus image classification. Despite being trained exclusively on synthetic data, RaSD exhibits robust transferability across retinal datasets, enabling consistent performance gains without relying on large-scale real-image pre-training. 

\begin{figure*}
    \centering
    \includegraphics[width=1.0\linewidth]{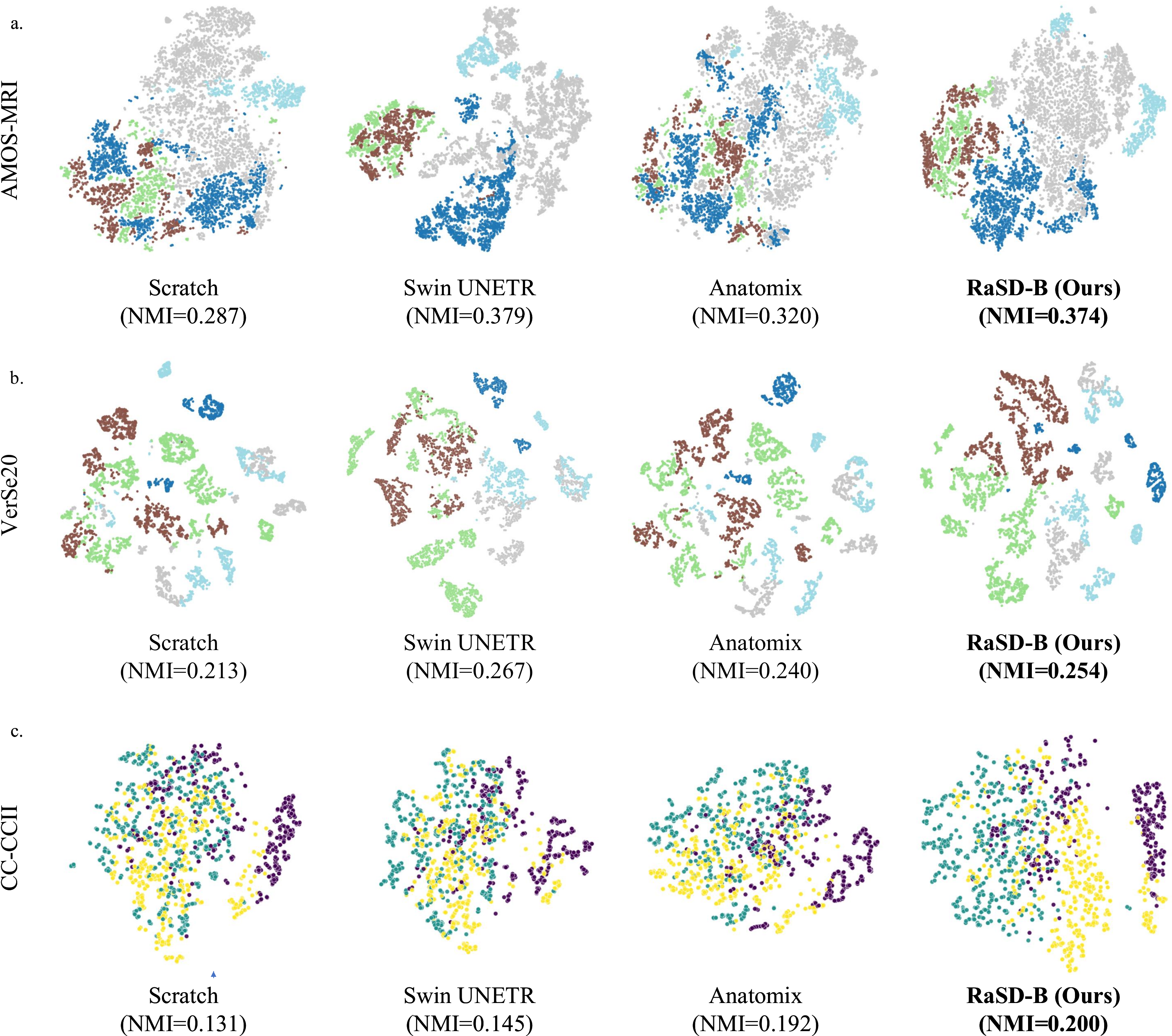}
    \caption{The t-SNE visualization of learned representations on AMOS-MR, VerSe20 and CC-CCII datasets, which demonstrates that RaSD produces more semantically coherent clusters.}
    \label{fig:model_analysis}
\end{figure*}

\subsection{RaSD delivers superior pathology segmentation}
RaSD-2D demonstrates strong and generalizable performance across a diverse spectrum of pathology segmentation tasks (Fig.~\ref{fig:Comparison3} h-p). To comprehensively evaluate our approach, we benchmarked it against four key baselines, including SOTA general-purpose foundation model LVM-Med (R50), LVM-Med (ViT), and MedSAM and a SOTA model specialized for pathology, PathSegmentator\cite{chen2025segment}. Our RaSD outperforms all general-purpose foundation models (LVM-Med R50/ViT and MedSAM) on four out of nine datasets (i, l, n, p), convincingly validating its superior representation learning capability. Our RaSD also achieves the highest Dice scores on the CoCaHis (i) and Janowczyk (l) datasets, surpassing the pathology-specific model PathSegmentor by 5.3\% and 8.1\% respectively. This demonstrates that our synthetic data pre-training RaSD is not only comparative with real-image pre-trained models but can even exceed the performance of modality-specific models in some tasks. These results show that learning through randomized synthesis and disentanglement provides a powerful and scalable pathway toward building high-performing foundation models without the dependency on large-scale real data.

\subsection{Pre-training and adaptation efficiency}
Our RaSD significantly enhances training efficiency through its online data generation capability, which effectively reduces storage overhead and accelerates the training process. For the online training paradigm, RaSD generates synthetic data during the training process. For data preparation, this approach only involves the immediate generation step without persistent storage, resulting in remarkably low preparation times. During model training, the synthesized data can be directly fed into the training pipeline, eliminating data I/O time and enabling efficient memory-to-memory data transfer. In contrast, the offline training paradigm requires complete data synthesis and storage before training. This two-step process incurs substantial overhead in data preparation, as it must both generate and persistently save all synthetic data. Furthermore, during the training phase, the offline approach must reload the pre-saved data from storage, introducing additional I/O latency that slows down the overall training process. As shown in Table~\ref{tab:online_comparison}, the online synthesis paradigm consistently outperforms its offline counterpart, achieving notable reductions in both data preparation and model training times across all tested data scales. At the scale of 1000 images, the online method completes the workflow 96.27 seconds faster, accounting for a significant 13\% overall improvement. Moreover, the storage advantage of RaSD’s online generation is particularly significant. While offline synthesis requires considerable storage capacity, our online approach maintains zero storage demand throughout the training workflow, which becomes increasingly beneficial as data scales grow.

Our RaSD model learns higher-quality and discriminative representations, as shown in Fig.~\ref{fig:model_analysis}. We employed t-distributed Stochastic Neighbor Embedding (t-SNE) \cite{maaten2008visualizing} to project the high-dimensional feature vectors extracted from the AMOS-MR, VerSe20, and CC-CCII datasets into a 2D space. The model trained from scratch yields a disorganized feature distribution, with points from different classes intermingled with each other. In comparison, synthetic-data-pretrained Anatomix provides a discernible but limited improvement, resulting in slightly more structured yet still overlapping clusters. The real-data-pretrained Swin UNETR demonstrates substantially better representation learning, forming relatively distinct clusters, particularly on the AMOS-MR dataset. Our RaSD method produces visibly more compact and better-separated clusters than the from-scratch baseline, while achieving performance comparable to real-image pre-trained models. This well-structured feature space provides qualitative evidence that RaSD learns highly discriminative. 

\section{Discussion}
Medical image analysis has seen significant advancements with the rise of FMs, yet their development has been consistently hampered by the scarcity, cost, and privacy concerns of large-scale real-world datasets. In this study, we propose RaSD, a synthetic data-driven pre-training framework that leverages synthetic images to reduce the reliance on real datasets, supporting efficient and generalizable MIFM development. Our comprehensive evaluation across 56 downstream tasks and 6 modalities provides compelling evidence that a FM pre-trained exclusively on synthetically generated data can achieve performance on par with or even superior to models trained on vast real-world datasets.

Our results demonstrate that RaSD effectively copes with the core limitations of real-based MIFM development. By generating data online from a Gaussian distribution, we eliminate the immense costs associated with data acquisition, annotation, and storage. The learning of disentanglement for the regions within the synthetic images further enables the pre-trained model to have a robust and generalizable understanding of anatomical structures. The observed performance gains, particularly in complex tasks such as retinal vessel segmentation and cross-modal transfer, highlight that our synthetic data engine is not merely a substitute for real data but a powerful mechanism for learning fundamental visual representations. For instance, RaSD's superior performance on MRI tasks compared to models pre-trained on CT datasets underscores its ability to learn modality-agnostic semantic features. This cross-modal generalization is a critical advantage, as it suggests a more robust and adaptable model, less prone to the domain-specific biases that limit other approaches.

Beyond its technical performance, RaSD's reliance on synthetic data offers a transformative paradigm shift for medical AI. By eliminating the need for patient-specific data, we effectively mitigate the most significant privacy challenges, paving the way for a truly privacy-preserving and scalable foundation model. This approach facilitates a ``free lunch'' in medical AI—where high-quality, diverse data is available on-demand, without the logistical and ethical burdens of real-world data collection. The scalability of RaSD, which was able to pre-train models with up to 653M parameters on millions of synthetic images, further validates its potential to support the next generation of larger, more powerful foundation models.

Although our study has provided strong support for the efficacy of synthetic data in MIFMs, it still has some limitations, which point to several important avenues for future work. First, the current approach relies on a rule-based synthesis engine, while this method is highly controllable and effective, it is difficult to capture the full complexity and subtle pathological variations present in real clinical data. For example, current RaSD is unable to synthesize rare disease manifestations or intricate anatomical anomalies that are not captured by our predefined rules. Future work will explore integrating our rule-based approach with generative models to create a hybrid system that leverages the best of both worlds, i.e., the structured control of rules with the organic diversity of generative models. Second, although RaSD performs exceptionally well across many tasks, there are instances where its performance falls slightly short of models specifically pre-trained on massive, real-world datasets for a narrow domain (e.g., specialized mammography models). This gap is caused by the inherent differences between our synthetic data distribution and the specific data distribution of highly specialized real-world datasets. Further refinement of our synthesis rules and the incorporation of more complex anatomical and pathological patterns may help close this gap. Lastly, 
although RaSD has been extensively validated across multiple imaging modalities including CT, MRI, X-ray, ultrasound, fundus, and pathology images, further evaluation on real-world clinical datasets is required to fully confirm its robustness and generalizability in practical medical scenarios.

\section{Methods} \label{sec:method}
\begin{figure}
    \centering
    \includegraphics[width=\linewidth]{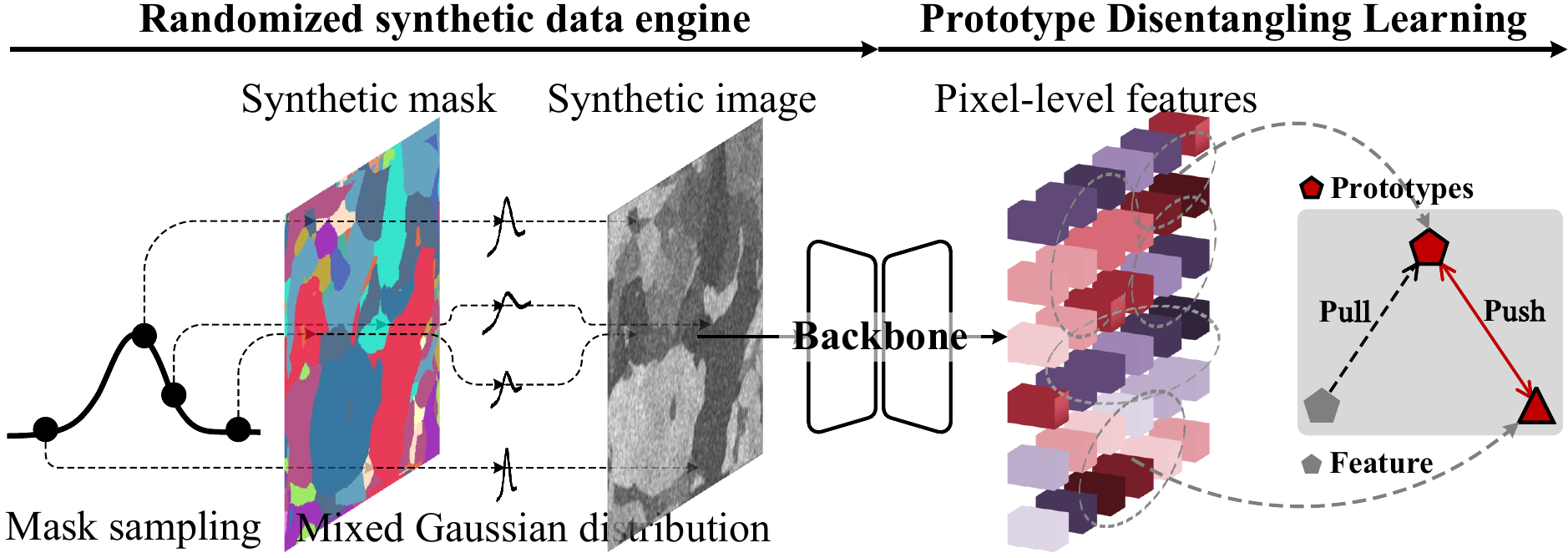}
    \caption{The overview of our RaSD framework. It synthesizes medical images using mixtures of Gaussian distributions to model anatomical structures and appearance variations, and trains the network to disentangle regions for discriminative representation.}
    \label{fig:overview}
\end{figure}
\begin{figure*}
    \centering
    \includegraphics[width=\linewidth]{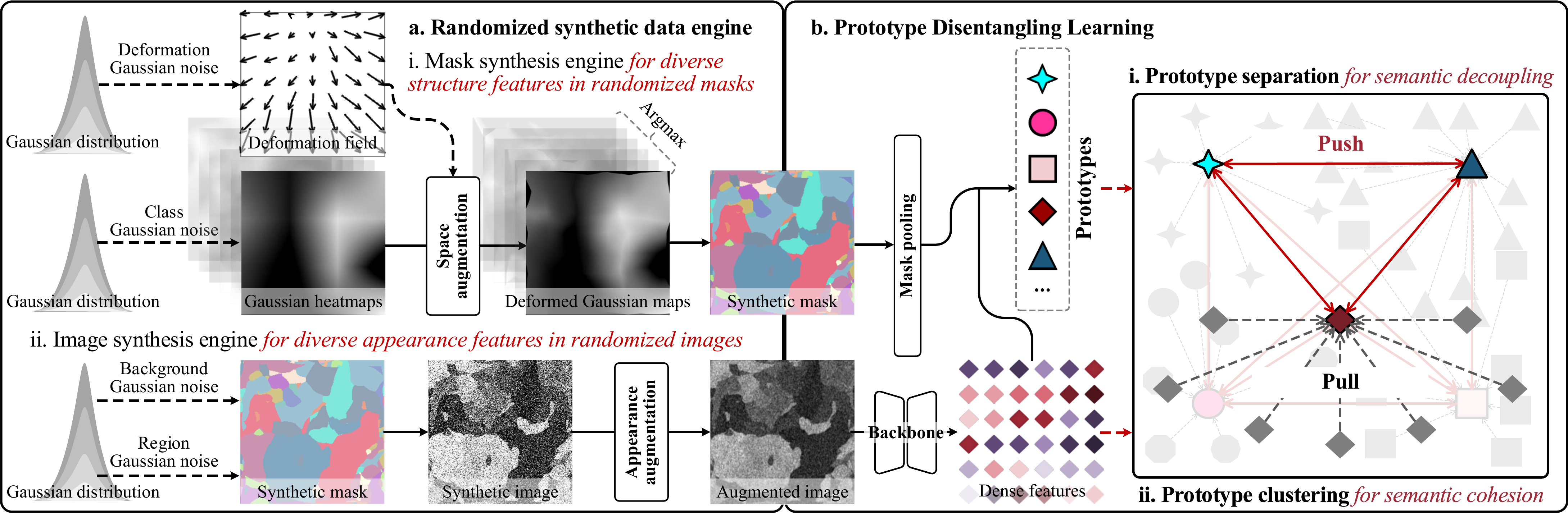}
    \caption{The detailed method of our RaSD for MIFM pre-training. a. Randomized synthetic data engine generates large-scale randomized medical image-mask pairs. It consists of a mask synthesis engine (i) for diverse structure features and an image synthesis engine (ii) for diverse appearance features. b. Prototype disentangling learning decouples complex semantics for effective semantic disentanglement ability in MIFMs. It learns prototype separation (i) and prototype clustering (ii) for the abilities of semantic decoupling and semantic cohesion.}
    \label{fig:MGDE}
\end{figure*}
Our RaSD pre-trains the MIFMs via synthetic medical images generated from Gaussian distributions and is able to achieve low-cost learning for transformable representation. As shown in Fig.~\ref{fig:overview}, it includes two key components. 1) Our randomized synthetic data engine generates synthetic images with pixel-level labels from Gaussian distributions, providing large-scale training data with diverse structures and appearances. 2) Our prototype disentangling learning learns to decouple complex semantics for effective semantic disentanglement ability in MIFMs.

\subsection{Randomized synthetic data engine}\label{sec:MGDE}
Our data engine generates large-scale synthetic randomized medical image-mask pairs using a mask synthesis engine and an image synthesis engine. These engines leverage class distributions under the assumption of a Gaussian distribution, subsequently synthesizing masks and images.

\textbf{Mask synthesis engine for diverse structure features:} Our mask synthesis engine (Fig.~\ref{fig:MGDE} a-i) simulates various objects in medical images from Gaussian distributions, reflecting their natural variability in geometry and density \cite{reynolds2009gaussian}. Specifically, it has two steps: 

\textit{a) Randomized sampling from Gaussian distributions.} It randomly samples Gaussian heatmaps $H=\{h_i\}_{i}^{I}$ that encode density distribution information and deformation field $\phi$ that encodes geometric  information. The Gaussian heatmaps are probabilities of $I$ classes randomly sampled from the Gaussian distribution. For each heatmap, we sample $N$ probability maps $\{p_n\}_{n}^{N}$ from the Gaussian distributions with different scales to simulate the anatomies with different scale granularities. These probability maps are unified to the same scale through linear interpolation $\mathcal{I}$ and summed to the final heatmap $p_i$. Similarly, the deformation field $\phi$ is also sampled from the Gaussian distribution in multiple scales with $d$ dimensions. This is formulated as
\begin{align}
    &\begin{matrix}H=\{\sum^{N}_{n}\mathcal{I}(p_n\sim\mathcal{N}(\mu,\sigma_{n}^2))\}^{I}\end{matrix},\\
    &\begin{matrix}\phi=\{\sum^{N}_{n}\mathcal{I}(\phi_{n}\sim\mathcal{N}(\mu,\sigma_{n}^2))\}^{d},\end{matrix}\notag
\end{align}
where the $\sigma_{n}$ is a random value from [0,1], and the $\mu$ is 0.

\textit{b) Deforming heatmaps to improve geometry diversity.} It utilizes the synthesized deformation field to warp the synthesized Gaussian heatmaps to increase the geometric diversity. It distorts the positions in the heatmaps according to the vectors in the deformation field via the space transformation \cite{hoffmann2021synthmorph}, thus producing a new structural form in the deformed Gaussian maps, i.e., $h'_i=\phi(h_i)$. Finally, the deformed maps are fused as the synthesis mask $y$ via $\arg\max_{i}(\{h'_i\}_{i}^{I})$. 

\textbf{Image synthesis engine for diverse appearance features:} Our image synthesis engine (Fig.~\ref{fig:MGDE} a-ii) simulates various imaging conditions and generates variations that reflect the natural variability of medical data, enhancing the diversity and realism of synthesized images in two steps:

\textit{a) Synthesizing region appearance to generate images.} For each class region $i$, it randomly samples the intensities from a Gaussian distribution with random mean $\mu$ and standard deviation (std) $\sigma$. Therefore, each class region will be filled with Gaussian noise with different means and stds for region appearance $r_i$. Background Gaussian noise $r_b$ is further added to the whole image to simulate the interference from the environment during imaging. This process is formulated as 
\begin{align}
    &\begin{matrix}r_i=\mathcal{N}(\mu_i,\sigma_{i}^2), r_b=\mathcal{N}(\mu_b,\sigma_{b}^2)\end{matrix},\\
    &\begin{matrix}x=r_b+\sum^{I}_{i}r_i,\end{matrix}
\end{align}
where $\mathcal{N}$ is the Gaussian distribution, $\mu_i,\mu_b$ are mean values randomly sampled from [0,1], $\sigma_i,\sigma_b$ are std values randomly sampled from [0,0.1], $x$ is the synthetic image.

\textit{b) Augmenting images to improve appearance diversity.} It simulates the image quality defects that will occur in medical imaging, thus synthesizing augmented images to improve its appearance diversity. It introduces a bias field \cite{song2017review} $B$ sampled from a Gaussian distribution to corrupt the image, enhancing the nonlinear effect of the whole image. Then, it applies Gaussian blur with a Gaussian kernel $K_{\sigma}$ on the whole image to enhance the resolution diversity of images and subsampling $\mathcal{S}$ on the Z axis to simulate the diversity of slice thickness \cite{billot2023synthseg} (for 3D situation). The augmentation process can be formulated as
\begin{equation}
    \hat{x}=\mathcal{S}(K*(xB),z),
\end{equation}
where the $z$ is the thickness on the Z axis, the $*$ is the matrix multiplication, and the $\hat{x}$ is the augmented image. This augmentation process accounts for factors that will alter pixel values in real data generation. Training a model to identify and restore labels affected by these transformations, the model will represent the properties of medical images.

\subsection{Prototype disentangling learning}
Our disentangling learning decouples complex semantics for effective semantic disentanglement ability in MIFMs. It reduces the coupling of inter-region features and increases the cohesion of inner-region features in a prototype learning from synthesized images and mask pairs, enabling the model to have a transformable semantic disentanglement ability.

\textbf{Overall learning framework:} As shown in Fig.~\ref{fig:MGDE} b, the synthetic images $\hat{x}$ are put into a backbone model $\mathcal{M}_{\theta}$ ($\theta$ is the parameter) with an encoder-decoder structure, e.g., Swin-UNETR \cite{hatamizadeh2021swin}, to extract dense features $f=\mathcal{M}_{\theta}(\hat{x})$. These features are grouped according to the corresponding regions on the synthetic mask $y$, and the features in each group are averaged \cite{wang2019panet} for $I$ prototypes $\textbf{\text{P}}=\{P_i\}_i^I$, i.e.,
\begin{equation}
    \begin{matrix}P_i = \frac{\sum_{j\in \{y_i\}}f_{j}}{\sum_{j\in \{y_i\}}1}\end{matrix}.
\end{equation}
These prototypes are further used to calculate prototype separation loss $\mathcal{L}_{ps}$ and prototype clustering loss $\mathcal{L}_{pc}$ to drive the model to learn the disentanglement ability.

\textbf{Prototype separation for semantic decoupling:} The prototype separation loss $\mathcal{L}_{ps}$ is calculated between the prototypes to reduce the coupling of the represented semantics. It enlarges the distance (scaled dot-product similarity) between prototypes by InfoNCE loss \cite{chen2020simple} to increase their discrimination. Therefore, the overall feature distribution representing the semantics of each region will be separated, thus improving the discrimination of the model, i.e.,
\begin{equation}
    \begin{matrix}\mathcal{L}_{ps}(\textbf{\text{P}})=-\sum_{i}^{I}\log\frac{\exp({P_iP_i^T}/\tau)}{\sum_{j}^{I}\exp({P_iP_j^T}/\tau)}\end{matrix},
\end{equation}
where $\tau$ is the temperature coefficient to control the separation degree, we set 0.07 in our experiment.

\textbf{Prototype clustering for semantic cohesion:} The prototype clustering loss $\mathcal{L}_{pc}$ is calculated between the prototypes $\textbf{\text{P}}$ and the dense features $f$ to improve cohesion of the features in the same semantic regions. Scaled dot-product similarity \cite{chen2020simple} is used to measure the distances between each prototype and the dense features and the distances belonging to the same classes $y$ are enlarged, i.e.,
\begin{equation}
    \begin{matrix}\mathcal{L}_{pc}(\textbf{\text{P}},f,y)=-\sum_{i}^{I}y_i\log\frac{\exp({P_i f^T}/\tau)}{\sum_{j}^{I}\exp({P_j f^T}/\tau)}\end{matrix}.
\end{equation}
Our proposed prototype disentangling learning (PDL) is the sum of the prototype separation and clustering losses, i.e., $\mathcal{L}_{PDL}=\mathcal{L}_{pc}+\mathcal{L}_{ps}$. Its dynamics can be formulated as $\theta^{*}\leftarrow\theta-\eta\frac{\partial\mathcal{L}_{PDL}(\theta,x,y)}{\partial\theta}$, where the $\eta$ is the learning rate. By optimizing the PDL loss, the model will be able to disentangle the semantics within medical images, enhancing its robustness in capturing complex medical image scene information. 

\subsection{Pre-training and adaptation} 

\textbf{Pre-training.} For the pre-training phase, our models were trained for $6 \times 10^5$ iterations. A learning rate of $1 \times 10^{-5}$ was used with a batch size of 2 for RaSD-3D, and 16 for RaSD-2D. In each iteration, a unique synthetic image-mask pair was generated. This on-the-fly synthesis eliminates the need for data storage and avoids I/O bottlenecks. Given that the entire synthesis process is a GPU-based matrix operation, it maintains strong real-time performance (0.05s/volume). Our 3D synthetic data had a size of $128 \times 128 \times 128$ and consisted of 8 distinct classes, chosen to represent the major anatomical regions while keeping the synthetic generation tractable. The 2D synthetic data was $512 \times 512$. All synthetic images were in grayscale.

\textbf{Adaptation.} For the fine-tuning phase, we follow the settings in prior work \cite{wu2024voco}. We used a learning rate of $3 \times 10^{-4}$ and optimized the models using the AdamW optimizer with a weight decay of $1 \times 10^{-5}$ and a momentum of 0.99. We applied standard data augmentation and pre-processing techniques, including scaling, shifting, flipping, and normalization. The best checkpoint was selected based on validation performance every $1 \times 10^4$ iterations.

\textbf{Implementation details.} Pre-training was conducted on NVIDIA L20 GPUs with 48 GB of memory, while downstream experiments were performed on NVIDIA RTX 4090D GPUs with 24 GB of memory. The framework was implemented using PyTorch \cite{paszke2019pytorch} and MONAI \cite{cardoso2022monai}, which provided modular and efficient support for medical image processing and model development.

\subsection{Evaluation metrics}
We report Dice similarity coefficient (Dice) for segmentation tasks across CT, MR, radiography, mammography, and fundus photography. For detection tasks, including mammography lesion detection, we additionally report the intersection over union (IoU) to evaluate localization performance. For classification tasks, such as disease diagnosis, fracture classification, breast cancer diagnosis, and retinal disease grading, we report AUC. 

\subsection{Comparison methods} 
We perform in-depth comparisons with existing pretrained models whose codes and checkpoints are publicly available. 
For 2D tasks, the compared methods include LVM-Med~\cite{mh2023lvm} and MedSAM~\cite{ma2024segment}. LVM-Med is evaluated with both ResNet-50 (R50) and Vision Transformer (ViT) backbones. In addition, we include modality-specific models for certain tasks: Mammo-CLIP~\cite{ghosh2024mammo} and MAMA~\cite{du2025multi} for mammography, and PathSegmentor~\cite{chen2025segment} for pathology. 
For 3D tasks, the compared methods include SwinUNETR~\cite{tang2022self}, SuPreM~\cite{li2024well}, VoCo~\cite{wu2024voco, wu2024large}, and Anatomix~\cite{dey2025learning}. Anatomix is also pretrained on synthetic data, making it particularly relevant for assessing the effectiveness of synthetic-data-based pretraining strategies.
All comparison methods are initialized with their officially released pretrained weights and subsequently fine-tuned on the target datasets following the same experimental protocols.

\section{Ethics Declaration}
This project was reviewed and approved by the Human and Artefacts Research Ethics Committee (HAREC) under protocol number HREP-2024-0429. All experiments were conducted in accordance with this approved protocol.

\section{Data Availability}
Detailed information on all datasets used in this study is provided in Appendix~\ref{sec:appendix_downstream}. All public datasets are available from their original sources, with direct links summarized in Table~\ref{tab:DataLink} for verification and further analysis.

\section{Code Availability}
The code developed for this study, including implementations of RaSD and scripts for reproducing the main experiments, is publicly available at https://github.com/yweibs/RaSD.

\section{Acknowledgment}
This work was supported by the National Natural Science Foundation of China (No. 62202403), the Innovation and Technology Commission (Project No. MHP/002/22, GHP/006/22GD and ITCPD/17-9), Research Grants Council of the Hong Kong Special Administrative Region, China (Project No: T45-401/22-N) and National Key R\&D Program of China (Project No. 2023YFE0204000).

\bibliography{egbib}

\begin{table*}
\centering
\caption{The links of publicly available datasets utilised in this study are detailed below.}
\makebox[\textwidth][c]{%
\begin{tabular}{ll}
\Xhline{1pt}
\textbf{Dataset}
& \textbf{Link}
\\
\hline
TotalSegmentator 
& \url{https://zenodo.org/records/14710732}
\\
BHSD
& \url{https://huggingface.co/datasets/Wendy-Fly/BHSD}
\\
BTCV
& \url{https://www.synapse.org/Synapse:syn3193805/wiki/217754}
\\
IRCADb
& \url{https://www.ircad.fr/research/data-sets/liver-segmentation-3d-ircadb-01/}
\\
CHAOS
& \url{https://chaos.grand-challenge.org/Publications/}
\\
MSD
& \url{http://medicaldecathlon.com/dataaws/index.html}
\\
Covid-19-20
& \url{https://covid-segmentation.grand-challenge.org/}
\\
Parse22
& \url{https://parse2022.grand-challenge.org/}
\\
SegThor
& \url{https://competitions.codalab.org/competitions/21145}
\\
MM-WHS
& \url{https://zmiclab.github.io/zxh/0/mmwhs/}
\\
VerSe20
& \url{https://verse2020.grand-challenge.org/}
\\
CC-CCII
& \url{http://ncov-ai.big.ac.cn/download?lang=en}
\\
LUNA16
& \url{https://luna16.grand-challenge.org/Data/}
\\
BraTS21
& \url{https://www.cancerimagingarchive.net/analysis-result/rsna-asnr-miccai-brats-2021}
\\
AMOS-MR
& \url{https://amos22.grand-challenge.org/}
\\
ATLAS-MR
& \url{https://atlas-challenge.u-bourgogne.fr/dataset}
\\
ACDC
& \url{https://www.creatis.insa-lyon.fr/Challenge/acdc/databasesTraining.html}
\\
MRNet
& \url{https://stanfordmlgroup.github.io/competitions/mrnet/}
\\
Shenzhen set
& \url{ http://archive.nlm.nih.gov/repos/chestImages.php}
\\
MC set
& \url{http://archive.nlm.nih.gov/repos/chestImages.php}
\\
RSNA-PBA
& \url{https://www.kaggle.com/datasets/kmader/rsna-bone-age}
\\
BFMR
& \url{https://www.kaggle.com/datasets/bmadushanirodrigo/fracture-multi-region-x-ray-data}
\\
KAU-BCMD
& \url{https://www.kaggle.com/datasets/asmaasaad/king-abdulaziz-university-mammogram-dataset}
\\
BMCD
& \url{https://zenodo.org/records/5036062}
\\
LAMIS-DMDB
& \url{https://github.com/LAMISDMDB/LAMISDMDB_Sample}
\\
CDD-CESM
& \url{https://www.cancerimagingarchive.net/collection/cdd-cesm/}
\\
MM
& \url{https://data.mendeley.com/datasets/fvjhtskg93/1}
\\
INbreast
& \url{https://www.kaggle.com/datasets/tommyngx/inbreast2012}
\\
VinDr-Mammo
& \url{https://vindr.ai/datasets/mammo}
\\
CBIS-DDSM
& \url{https://www.cancerimagingarchive.net/collection/cbis-ddsm/}
\\
STARE
& \url{https://cecas.clemson.edu/~ahoover/stare/probing/index.html}
\\
DRIVE
& \url{https://drive.grand-challenge.org/DRIVE/}
\\
HVDROPDB
& \url{https://data.mendeley.com/datasets/xw5xc7xrmp/3}
\\
IDRiD
& \url{https://idrid.grand-challenge.org/}
\\
DRAC
& \url{https://drac22.grand-challenge.org/}
\\
GlaS
& \url{https://warwick.ac.uk/fac/cross_fac/tia/data/glascontest/}
\\
CoCaHis
& \url{https://cocahis.irb.hr/}
\\
TNBC
& \url{https://zenodo.org/records/1175282}
\\
MoNuSAC
& \url{https://monusac-2020.grand-challenge.org/}
\\
Janowczyk
& \url{https://andrewjanowczyk.com/deep-learning/}
\\
CRAG
& \url{https://warwick.ac.uk/fac/sci/dcs/research/tia/data/mildnet}
\\
MIDOG
& \url{https://imig.science/midog2021/download-dataset/}
\\
NuCLS
& \url{https://nucls.grand-challenge.org/}
\\
WSSS4LUAD
& \url{https://wsss4luad.grand-challenge.org/}
\\
TN3K
& \url{https://github.com/haifangong/TRFE-Net-for-thyroid-nodule-segmentation}
\\
TG3K
& \url{https://github.com/haifangong/TRFE-Net-for-thyroid-nodule-segmentation}
\\
DDTI
& \url{http://cimalab.intec.co/applications/thyroid}
\\
MMOTU
& \url{https://github.com/cv516Buaa/MMOTU_DS2Net}
\\
\Xhline{1pt}
\end{tabular}
}
\label{tab:DataLink}
\end{table*}

\begin{table*}
\centering
\caption{Summary of medical CT datasets used in this study. The collection encompasses diverse anatomical regions and clinical tasks.}
\small
\begin{tabular}{l l l c cc}
\toprule
\multirow{2}{*}{\textbf{Dataset}} 
& \multirow{2}{*}{\textbf{Task}} 
& \multirow{2}{*}{\textbf{Anatomical Region}} 
& \multirow{2}{*}{\textbf{\#Classes}} 
& \multicolumn{2}{c}{\textbf{\#Volumes}} \\
\cmidrule(lr){5-6}
& & & & \textbf{Train} & \textbf{Valid} \\
\midrule
TotalSegmentator~\cite{wasserthal2023totalsegmentator}
& Segmentation & Whole Body & 104 & 911 & 228 \\

BHSD~\cite{wu2023bhsd}
& Segmentation & Brain Bleed & 5 & 153 & 39 \\

BTCV~\cite{landman2015miccai}
& Segmentation & Abdomen & 13 & 24 & 6 \\

IRCADb~\cite{soler20103d}
& Segmentation & Liver Tumor & 2 & 15 & 5 \\

CHAOS~\cite{kavur2021chaos}
& Segmentation & Liver & 1 & 16 & 4 \\

MSD07~\cite{antonelli2022medical}
& Segmentation & Pancreas Tumor & 1 & 257 & 24 \\

MSD09~\cite{antonelli2022medical}
& Segmentation & Spleen & 1 & 32 & 9 \\

Covid-19-20~\cite{roth2022rapid}
& Segmentation & Covid & 1 & 159 & 40 \\

Parse22~\cite{luo2023efficient}
& Segmentation & Pulmonary Artery & 1 & 80 & 20 \\

SegThor~\cite{lambert2020segthor}
& Segmentation & Thoracic Organs & 4 & 32 & 8 \\

WHS-CT~\cite{zhuang2018multivariate}
& Segmentation & Heart & 7 & 16 & 4 \\

VerSe20~\cite{sekuboyina2021verse}
& Segmentation & Vertebrae & 28 & 99 & 8 \\

\midrule
CC-CCII~\cite{zhang2020clinically}
& Classification & Covid & 3 & 2516 & 1664 \\

LUNA16~\cite{setio2017validation}
& Classification & Lung Nodule & 2 & 445 & 178 \\
\bottomrule
\end{tabular}
\label{tab:CTDatasets}
\end{table*}

\newpage

\section{Appendix}

\subsection{Downstream Tasks} \label{sec:appendix_downstream}

\begin{table*}
\centering
\small 
\caption{Comparison of CT segmentation performance. The table presents Dice Similarity Coefficient (DSC) results across 12 public datasets. The columns `p vs Scratch' and `p vs Best' report the p-values of our RaSD compared to the from-scratch method (Swin Scratch) and the best-performing method, respectively.}
\resizebox{\linewidth}{!}{%
\begin{tabular}{lcccccccc}
\toprule
\textbf{Dataset} 
& \textbf{Swin (Scratch)} 
& \textbf{Swin-UNETR} 
& \textbf{SuPreM} 
& \textbf{VoCo (160k)} 
& \textbf{Anatomix} 
& \textbf{RaSD-3D (Ours)} 
& \textbf{p vs Scratch}
& \textbf{p vs Best} \\
\midrule
TotalSegmentator
& \shortstack{75.6 \\ (74.2--77.0)} 
& \shortstack{79.7 \\ (78.4--81.0)} 
& \shortstack{81.2 \\ (80.0--82.4)} 
& \shortstack{86.2 \\ (85.1--87.3)} 
& \shortstack{81.5 \\ (80.3--82.7)} 
& \shortstack{\textbf{81.7} \\ (\textbf{80.5--82.9})} 
& <0.01
& <0.01\\
\midrule
BHSD
& \shortstack{34.4 \\ (26.2--42.5)} 
& \shortstack{39.4 \\ (31.7--41.7)} 
& \shortstack{39.5 \\ (31.6--47.3)} 
& \shortstack{39.8 \\ (31.8--47.4)} 
& \shortstack{ 36.9\\ (29.0--44.8)} 
& \shortstack{\textbf{39.0} \\ (\textbf{30.8--47.1})} 
& 0.01
& 0.59\\
\midrule
BTCV
& \shortstack{81.5 \\ (80.9--82.9)} 
& \shortstack{86.4 \\ (83.9--88.8)}
& \shortstack{87.3 \\ (85.4--89.1)} 
& \shortstack{86.5 \\ (85.5--87.5)}
& \shortstack{84.2 \\ (83.2--85.1)}
& \shortstack{\textbf{84.6} \\ (\textbf{83.6--85.6})} 
& <0.01
& 0.01\\
\midrule
IRCADb
& \shortstack{57.2 \\ (31.6--82.8)} 
& \shortstack{68.0 \\ (49.8--86.1)} 
& \shortstack{68.5 \\ (53.1--83.8)} 
& \shortstack{74.3 \\ (62.8--85.7)} 
& \shortstack{67.0 \\ (52.8--81.2)} 
& \shortstack{\textbf{74.3} \\ (\textbf{60.2--88.5})} 
& 0.18
& 0.98\\
\midrule
CHAOS
& \shortstack{94.5 \\ (92.2--96.8)} 
& \shortstack{96.1 \\ (94.9--97.3)} 
& \shortstack{96.5 \\ (95.5--97.5)} 
& \shortstack{96.4 \\ (95.4--97.4)} 
& \shortstack{93.8 \\ (89.5--98.0)} 
& \shortstack{\textbf{95.5} \\ (\textbf{94.4--96.6})} 
& 0.21
& 0.13\\
\midrule
MSD09
& \shortstack{94.9 \\ (92.6--97.3)} 
& \shortstack{94.8 \\ (93.2--96.3)} 
& \shortstack{95.9 \\ (94.6--97.2)} 
& \shortstack{96.2 \\ (95.1--97.2)} 
& \shortstack{95.1 \\ (93.4--96.8)} 
& \shortstack{\textbf{95.3} \\ (\textbf{93.7--96.9})} 
& 0.53
& 0.09\\
\midrule
MSD07
& \shortstack{73.0 \\ (67.1--77.5)} 
& \shortstack{71.9 \\ (66.6--77.1)} 
& \shortstack{74.6 \\ (69.6--79.6)} 
& \shortstack{74.7 \\ (70.3--79.1)} 
& \shortstack{73.0 \\ (68.0--78.1)} 
& \shortstack{\textbf{73.6} \\ (\textbf{68.2--79.0})} 
& 0.07
& 0.46\\
\midrule
Covid-19-20
& \shortstack{67.8 \\ (60.5--75.1)} 
& \shortstack{68.1 \\ (61.5--74.7)} 
& \shortstack{63.3 \\ (55.5--71.1)} 
& \shortstack{68.8 \\ (61.1--76.4)} 
& \shortstack{66.9 \\ (59.4--74.4)} 
& \shortstack{\textbf{68.9} \\ (\textbf{61.3--76.5})} 
& 0.22
& 0.92\\
\midrule
Parse22
& \shortstack{82.8 \\ (80.6--85.1)} 
& \shortstack{83.5 \\ (81.6--85.4)} 
& \shortstack{83.9 \\ (82.1--85.7)} 
& \shortstack{82.8 \\ (80.6--84.9)} 
& \shortstack{83.0 \\ (80.9--85.2)} 
& \shortstack{\textbf{83.8} \\ (\textbf{81.7--85.9})} 
& <0.01
& 0.75\\
\midrule
SegThor
& \shortstack{89.2 \\ (86.8--91.7)} 
& \shortstack{89.2 \\ (86.7--91.8)} 
& \shortstack{89.9 \\ (87.7--92.2)} 
& \shortstack{90.0 \\ (87.7--92.2)} 
& \shortstack{89.3 \\ (87.0--91.6)} 
& \shortstack{\textbf{89.8} \\ (\textbf{87.1--92.4})} 
& 0.08
& 0.60\\
\midrule
WHS-CT
& \shortstack{90.5 \\ (87.8--93.3)} 
& \shortstack{90.1 \\ (86.8--93.3)} 
& \shortstack{90.7 \\ (87.9--93.5)} 
& \shortstack{90.8 \\ (87.9--93.8)} 
& \shortstack{90.3 \\ (87.3--93.2)} 
& \shortstack{\textbf{91.0} \\ (\textbf{88.0--94.0})} 
& 0.06
& 0.03\\
\midrule
VerSe20
& \shortstack{76.4 \\ (65.6--87.1)} 
& \shortstack{81.7 \\ (74.1--89.3)} 
& \shortstack{87.3 \\ (81.5--93.2)} 
& \shortstack{85.5 \\ (78.4--92.6)} 
& \shortstack{80.2 \\ (67.1--93.4)} 
& \shortstack{\textbf{84.0} \\ (\textbf{71.3--96.7})} 
& 0.11
& 0.48\\
\bottomrule
\end{tabular}%
}
\label{tab:CT_seg}
\end{table*}

\begin{table*}
\centering
\small 
\caption{Comparison of classification performance for CT and MR datasets. The table presents results across 3 datasets, measured by AUC in percentage.}
\resizebox{\linewidth}{!}{%
\begin{tabular}{lcccccccc}
\toprule
\textbf{Dataset} 
& \textbf{Swin (Scratch)} 
& \textbf{Swin-UNETR} 
& \textbf{SuPreM} 
& \textbf{VoCo (160k)} 
& \textbf{Anatomix} 
& \textbf{RaSD-3D (Ours)} 
& \textbf{p vs Scratch}
& \textbf{p vs Best} \\
\midrule
CC-CCII
& \shortstack{98.0 \\ (97.2--98.8)} 
& \shortstack{98.3 \\ (97.4--99.3)} 
& \shortstack{98.3 \\ (97.5--99.1)} 
& \shortstack{98.8 \\ (98.5--99.1)} 
& \shortstack{97.8 \\ (97.2--98.5)} 
& \shortstack{\textbf{98.8} \\ (\textbf{98.2--99.4})} 
& <0.01
& 0.50 \\
\midrule
LUNA16
& \shortstack{96.5 \\ (96.5--96.5)} 
& \shortstack{97.6 \\ (97.6--97.6)} 
& \shortstack{97.9 \\ (97.8--97.9)} 
& \shortstack{98.7 \\ (98.7--98.7)} 
& \shortstack{97.9 \\ (97.9--97.9)} 
& \shortstack{\textbf{98.1} \\ (\textbf{98.0--98.1})} 
& <0.01
& <0.01\\
\midrule
MRNet
& \shortstack{79.7 \\ (70.7--89.2)} 
& \shortstack{82.6 \\ (73.3--91.9)} 
& \shortstack{82.8 \\ (70.6--95.1)} 
& \shortstack{81.4 \\ (71.2--91.6)} 
& \shortstack{81.4 \\ (71.1--81.8)} 
& \shortstack{\textbf{80.9} \\ (\textbf{70.8--91.0})} 
& 0.89
& 0.99\\
\bottomrule
\end{tabular}%
}
\label{tab:CT_MR_Classification}
\end{table*}

\begin{table*}
\centering
\caption{Summary of medical MR datasets used in this study. The collection encompasses diverse anatomical regions and clinical tasks.}
\small
\begin{tabular}{l l l c cc}
\toprule
\multirow{2}{*}{\textbf{Dataset}} 
& \multirow{2}{*}{\textbf{Task}} 
& \multirow{2}{*}{\textbf{Anatomical Region}} 
& \multirow{2}{*}{\textbf{\#Classes}} 
& \multicolumn{2}{c}{\textbf{\#Volumes}} \\
\cmidrule(lr){5-6}
& & & & \textbf{Train} & \textbf{Valid} \\
\midrule
BraTS21~\cite{baid2021rsna}
& Segmentation & Brain Tumor & 3 & 1000 & 251 \\

AMOS-MR~\cite{ji2022amos}
& Segmentation & Abdomen & 15 & 48 & 12 \\

ATLAS-MR~\cite{quinton2023tumour}
& Segmentation & Liver Tumor & 2 & 45 & 15 \\

WHS-MR~\cite{zhuang2018multivariate}
& Segmentation & Heart & 7 & 15 & 5 \\

ACDC~\cite{bernard2018deep}
& Segmentation & Heart & 3 & 160 & 40 \\

\midrule
MRNet~\cite{bien2018deep}
& Classification & Knee & 2 & 1130 & 120 \\
\bottomrule
\end{tabular}
\label{tab:MRDatasets}
\end{table*}

\begin{table*}
\centering
\small 
\caption{Comparison of MR segmentation performance. The table presents Dice Similarity Coefficient (DSC) results across 5 public datasets.}
\resizebox{\linewidth}{!}{%
\begin{tabular}{lcccccccc}
\toprule
\textbf{Dataset} 
& \textbf{Swin (Scratch)} 
& \textbf{Swin-UNETR} 
& \textbf{SuPreM} 
& \textbf{VoCo (160k)} 
& \textbf{Anatomix} 
& \textbf{RaSD-3D (Ours)} 
& \textbf{p vs Scratch}
& \textbf{p vs Best} \\
\midrule
BraTS21
& \shortstack{89.9 \\ (88.9--91.0)} 
& \shortstack{90.3 \\ (89.4--91.3)} 
& \shortstack{90.2 \\ (89.1--91.2)} 
& \shortstack{90.2 \\ (89.2--91.2)} 
& \shortstack{90.0 \\ (88.9--91.0)} 
& \shortstack{\textbf{90.4} \\ (\textbf{89.4--91.3})} 
& <0.01
& 0.82\\
\midrule
AMOS-MR
& \shortstack{71.5 \\ (64.4--78.6)} 
& \shortstack{72.6 \\ (65.6--79.5)} 
& \shortstack{74.0 \\ (67.0--81.1)} 
& \shortstack{77.2 \\ (70.5--83.9)} 
& \shortstack{71.2 \\ (64.3--78.1)} 
& \shortstack{\textbf{76.0} \\ (\textbf{69.3--82.8})} 
& <0.01
& <0.01 \\
\midrule
ATLAS-MR
& \shortstack{58.8 \\ (52.0--65.7)} 
& \shortstack{61.4 \\ (54.6--68.2)} 
& \shortstack{66.2 \\ (59.2--73.2)} 
& \shortstack{67.5 \\ (59.8--75.2)} 
& \shortstack{62.2 \\ (56.0--68.5)} 
& \shortstack{\textbf{67.6} \\ (\textbf{62.3--73.0})} 
& <0.01
& 0.96\\
\midrule
WHS-MR
& \shortstack{86.5 \\ (85.0--87.9)} 
& \shortstack{87.8 \\ (85.7--89.9)} 
& \shortstack{87.9 \\ (85.7--90.1)} 
& \shortstack{87.7 \\ (85.4--90.0)} 
& \shortstack{87.2 \\ (85.4--88.9)} 
& \shortstack{\textbf{87.7} \\ (\textbf{85.3--90.0})} 
& 0.16
& 0.72\\
\midrule
ACDC
& \shortstack{87.9 \\ (86.8--89.0)} 
& \shortstack{88.8 \\ (87.7--89.9)} 
& \shortstack{89.1 \\ (88.1--90.1)} 
& \shortstack{89.5 \\ (88.6--90.5)} 
& \shortstack{89.6 \\ (88.6--90.7)} 
& \shortstack{\textbf{89.9} \\ (\textbf{89.0--90.8})} 
& <0.01
& 0.49\\
\bottomrule
\end{tabular}%
}
\label{tab:MR_Seg}
\end{table*}

\begin{table*}
\centering
\caption{Summary of medical X-ray datasets used in this study. The collection encompasses diverse anatomical regions and clinical tasks.}
\small
\begin{tabular}{l l l cc}
\toprule
\multirow{2}{*}{\textbf{Dataset}} 
& \multirow{2}{*}{\textbf{Task}} 
& \multirow{2}{*}{\textbf{Anatomical Region}}  
& \multicolumn{2}{c}{\textbf{\#Images}} \\
\cmidrule(lr){4-5}
& & & \textbf{Train} & \textbf{Valid} \\
\midrule
Shenzhen-Seg~\cite{jaeger2014two}
& Segmentation & Chest & 465 & 101 \\

MC-Seg~\cite{jaeger2014two}
& Segmentation & Chest & 110 & 28 \\

RSNA-PBA~\cite{halabi2019rsna}
& Segmentation & Bone & 112 & 81 \\

\midrule
Shenzhen-Cls~\cite{jaeger2014two}
& Classification & Chest & 602 & 60 \\

MC-Cls~\cite{jaeger2014two}
& Classification & Chest & 117 & 21 \\

BFMR~\cite{fracturexray}
& Classification & Bone & 9240 & 823 \\
\bottomrule
\end{tabular}
\label{tab:XrayDatasets}
\end{table*}

\begin{table*}[ht]
\centering
\caption{Summary of mammography datasets used in this study. The table details the clinical task, specific prediction target, and dataset splits.}
\small
\begin{tabular}{l l l ccc}
\toprule
\multirow{2}{*}{\textbf{Dataset}} 
& \multirow{2}{*}{\textbf{Task}} 
& \multirow{2}{*}{\textbf{Specific Target}} 
& \multicolumn{3}{c}{\textbf{\#Images}} \\
\cmidrule(lr){4-6}
& & & \textbf{Train} & \textbf{Valid} & \textbf{Test} \\
\midrule
KAU-BCMD~\cite{alsolami2021king}
& Classification & Composition & 1662 & 236 & 480 \\

BMCD~\cite{loizidou2021breast}
& Classification & Composition & 280 & 40 & 80 \\

LAMIS-DMDB~\cite{imane2024lamis}
& Classification & BI-RADS & 1548 & 221 & 443 \\

BMCD~\cite{loizidou2021breast}
& Classification & BI-RADS & 280 & 40 & 80 \\

CDD-CESM~\cite{khaled2022categorized}
& Classification & Pathology & 463 & 66 & 133 \\

MM~\cite{aqdar2024mammogram}
& Classification & Pathology & 511 & 73 & 161 \\

\midrule
INbreast-Det~\cite{moreira2012inbreast}
& Detection & -- & 76 & 12 & 19 \\

VinDr-Mammo~\cite{nguyen2023vindr}
& Detection & -- & 802 & 113 & 198 \\

\midrule
CBIS-DDSM~\cite{lee2017curated}
& Segmentation & -- & 1075 & 142 & 344 \\

INbreast-Seg~\cite{moreira2012inbreast}
& Segmentation & -- & 76 & 12 & 19 \\
\bottomrule
\end{tabular}
\label{tab:MammoDatasets}
\end{table*}

\begin{table*}
\centering
\small 
\caption{Comparison of segmentation performance for chest X-ray datasets. The table presents results for the Shenzhen, Montgomery County (MC), and RSNA-PBA datasets, measured by Dice Similarity Coefficient (DSC) in percentage.}
\resizebox{\linewidth}{!}{%
\begin{tabular}{lcccccccc}
\toprule
\textbf{Dataset} 
& \textbf{UNet (Scratch)} 
& \textbf{LVM-Med (R50)} 
& \textbf{LVM-Med (ViT)} 
& \textbf{MedSAM} 
& \textbf{RaSD-2D (Ours)} 
& \textbf{p vs Scratch}
& \textbf{p vs Best} \\
\midrule
Shenzhen set
& \shortstack{81.2 \\ (79.7--82.6)} 
& \shortstack{76.6 \\ (74.2--79.0)} 
& \shortstack{89.3 \\ (88.6--90.0)} 
& \shortstack{87.9 \\ (87.1--88.7)} 
& \shortstack{\textbf{83.5} \\ (\textbf{82.7--84.4})} 
& <0.01
& <0.01\\
\midrule
MC set
& \shortstack{54.9 \\ (50.6--59.2)} 
& \shortstack{68.0 \\ (65.0--71.1)} 
& \shortstack{84.2 \\ (82.1--86.4)} 
& \shortstack{84.6 \\ (82.6--86.6)} 
& \shortstack{\textbf{62.0} \\ (\textbf{58.1--65.9})} 
& <0.01
& <0.01\\
\midrule
RSNA-PBA
& \shortstack{95.3 \\ (95.1--95.4)} 
& \shortstack{99.1 \\ (99.1--99.2)} 
& \shortstack{99.7 \\ (99.6--99.8)} 
& \shortstack{98.9 \\ (98.7--99.2)} 
& \shortstack{\textbf{99.7} \\ (\textbf{99.6--99.7})} 
& <0.01
& 0.91\\
\bottomrule
\end{tabular}%
}
\label{tab:Xray_Seg}
\end{table*}

\begin{table*}
\centering
\small 
\caption{Comparison of classification performance for X-ray datasets. The table presents results for the Shenzhen, Montgomery County (MC), and BFMR datasets, measured by AUC in percentage.}
\resizebox{\linewidth}{!}{%
\begin{tabular}{lccccccc}
\toprule
\textbf{Dataset} 
& \textbf{UNet (Scratch)} 
& \textbf{LVM-Med (R50)} 
& \textbf{LVM-Med (ViT)} 
& \textbf{MedSAM} 
& \textbf{RaSD-2D (Ours)} 
& \textbf{p vs Scratch}
& \textbf{p vs Best} \\
\midrule
Shenzhen set
& \shortstack{83.1 \\ (80.7-85.6)} 
& \shortstack{91.6 \\ (89.7--93.4)} 
& \shortstack{88.2 \\ (86.1--90.3)} 
& \shortstack{86.3 \\ (84.1--88.6)} 
& \shortstack{\textbf{85.9} \\ (\textbf{83.6--88.2})} 
& <0.01
& <0.01\\
\midrule
MC set
& \shortstack{70.9 \\ (62.4--79.4)} 
& \shortstack{73.6 \\ (65.4--81.9)} 
& \shortstack{86.4 \\ (80.0--92.8)} 
& \shortstack{79.1 \\ (71.5--86.7)} 
& \shortstack{\textbf{82.7} \\ (\textbf{75.7--89.8})} 
& <0.01
& <0.01\\
\midrule
BFMR
& \shortstack{90.2 \\ (90.0--90.0)} 
& \shortstack{96.9 \\ (96.8--96.9)} 
& \shortstack{99.9 \\ (99.9-100.0)} 
& \shortstack{99.9 \\ (99.8-99.9)} 
& \shortstack{\textbf{94.1} \\ (\textbf{94.0-94.0})} 
& <0.01
& <0.01\\
\bottomrule
\end{tabular}%
}
\label{tab:Xray_Cls}
\end{table*}

\begin{table*}
\centering
\tiny
\caption{Comparison of classification performance for mammography datasets. The table presents results on BI-RADS, composition, and pathology tasks across six public datasets, measured by AUC in percentage.}
\makebox[\textwidth][c]{%
\begin{tabular}{lccccccccc}
\toprule
\textbf{Dataset} 
& \textbf{UNet (Scratch)} 
& \textbf{LVM-Med (R50)} 
& \textbf{LVM-Med (ViT)} 
& \textbf{MedSAM} 
& \textbf{Mammo-CLIP} 
& \textbf{MAMA} 
& \textbf{RaSD-2D (Ours)} 
& \textbf{p vs Scratch}
& \textbf{p vs Best} \\
\midrule
\shortstack{KAU-BCMD \\ (BI-RADS)}
& \shortstack{74.4 \\ (67.1--81.2)} 
& \shortstack{89.6 \\ (87.2--91.8)} 
& \shortstack{88.3 \\ (84.8--91.5)} 
& \shortstack{81.5 \\ (77.4--86.0)} 
& \shortstack{93.5 \\ (91.3--95.3)} 
& \shortstack{93.4 \\ (90.9--95.6)} 
& \shortstack{\textbf{90.1} \\ (\textbf{87.7--92.1})} 
& <0.01 
& <0.01 \\
\midrule
\shortstack{BMCD \\ (BI-RADS)}
& \shortstack{53.3 \\ (41.1--64.9)} 
& \shortstack{80.7 \\ (74.2--86.9)} 
& \shortstack{74.4 \\ (67.7--80.6)}
& \shortstack{68.7 \\ (61.9--75.5)}
& \shortstack{89.9 \\ (84.4--94.4)} 
& \shortstack{89.9 \\ (83.1--93.6)} 
& \shortstack{\textbf{74.0} \\ (\textbf{67.1--80.3})} 
& <0.01 
& <0.01 \\
\midrule
\shortstack{LAMIS-DMDB \\ (Composition)}
& \shortstack{67.9 \\ (64.6--71.7)} 
& \shortstack{80.7 \\ (77.8--84.3)} 
& \shortstack{80.6 \\ (77.6--83.7)} 
& \shortstack{80.0 \\ (77.1--84.1)} 
& \shortstack{82.7 \\ (79.2--85.9)} 
& \shortstack{82.7 \\ (79.4--85.7)} 
& \shortstack{\textbf{80.0} \\ (\textbf{76.6--83.1})} 
& <0.01 
& <0.01 \\
\midrule
\shortstack{BMCD \\ (Composition)}
& \shortstack{56.3 \\ (46.5--67.4)} 
& \shortstack{90.0 \\ (84.2--94.2)} 
& \shortstack{94.0 \\ (90.0--97.2)} 
& \shortstack{88.5 \\ (83.4--93.1)} 
& \shortstack{95.8 \\ (92.7--98.3)} 
& \shortstack{97.5 \\ (95.4--99.2)} 
& \shortstack{\textbf{91.3} \\ (\textbf{86.1--95.5})} 
& <0.01 
& <0.01 \\
\midrule
\shortstack{CDD-CESM \\ (Pathology)}
& \shortstack{69.9 \\ (61.0--78.8)} 
& \shortstack{73.7 \\ (68.3--79.8)} 
& \shortstack{75.0 \\ (69.8--79.8)} 
& \shortstack{73.4 \\ (68.1--78.2)} 
& \shortstack{80.0 \\ (74.9--84.5)} 
& \shortstack{81.7 \\ (77.1--86.0)} 
& \shortstack{\textbf{78.5} \\ (\textbf{70.4--85.5})} 
& <0.01 
& <0.01 \\
\midrule
\shortstack{MM \\ (Pathology)}
& \shortstack{90.3 \\ (79.3--98.7)} 
& \shortstack{99.1 \\ (96.9--100.0)} 
& \shortstack{98.1 \\ (95.8-99.7)} 
& \shortstack{97.5 \\ (92.2--100.0)} 
& \shortstack{99.7 \\ (99.1-100.0)} 
& \shortstack{100.0 \\ (99.8-100.0)} 
& \shortstack{\textbf{99.2} \\ (\textbf{97.9-100.0})} 
& <0.01 
& <0.01 \\
\bottomrule
\end{tabular}%
}
\label{tab:Mammo_Cls}
\end{table*}

\begin{table*}
\centering
\tiny 
\caption{Comparison of detection performance for mammography datasets. The table presents results for the INbreast and VinDr-Mammo datasets, measured by Intersection over Union (IoU) in percentage.}
\makebox[\textwidth][c]{%
\begin{tabular}{lccccccccc}
\toprule
\textbf{Dataset} 
& \textbf{UNet (Scratch)} 
& \textbf{LVM-Med (R50)} 
& \textbf{LVM-Med (ViT)} 
& \textbf{MedSAM} 
& \textbf{Mammo-CLIP} 
& \textbf{MAMA} 
& \textbf{RaSD-2D (Ours)} 
& \textbf{p vs Scratch}
& \textbf{p vs Best} \\
\midrule
INbreast
& \shortstack{62.5 \\ (57.7--66.3)} 
& \shortstack{51.0 \\ (46.9--54.1)} 
& \shortstack{60.9 \\ (55.9--64.6)} 
& \shortstack{51.1 \\ (47.0--54.2)} 
& \shortstack{60.3 \\ (55.4--64.0)} 
& \shortstack{56.2 \\ (51.7--59.6)} 
& \shortstack{\textbf{66.6} \\ (\textbf{62.6--69.6})} 
& <0.01
& <0.01 \\
\midrule
VinDr-Mammo
& \shortstack{48.0 \\ (44.4--50.5)} 
& \shortstack{38.9 \\ (35.7--41.2)} 
& \shortstack{46.2 \\ (42.5--49.0)} 
& \shortstack{45.0 \\ (41.4--47.8)} 
& \shortstack{45.6 \\ (41.9--48.3)} 
& \shortstack{46.2 \\ (42.5--49.0)} 
& \shortstack{\textbf{49.4} \\ (\textbf{46.3--51.6})} 
& <0.01 
& <0.01 \\
\bottomrule
\end{tabular}%
}
\label{tab:Mammo_Det}
\end{table*}

\begin{table*}
\centering
\tiny 
\caption{Comparison of segmentation performance for mammography datasets. The table presents results for the CBIS-DDSM and INbreast datasets, measured by Dice Similarity Coefficient (DSC) in percentage.}
\makebox[\textwidth][c]{%
\begin{tabular}{lccccccccc}
\toprule
\textbf{Dataset} 
& \textbf{UNet (Scratch)} 
& \textbf{LVM-Med (R50)} 
& \textbf{LVM-Med (ViT)} 
& \textbf{MedSAM} 
& \textbf{Mammo-CLIP} 
& \textbf{MAMA} 
& \textbf{RaSD-2D (Ours)} 
& \textbf{p vs Scratch}
& \textbf{p vs Best} \\
\midrule
CBIS-DDSM
& \shortstack{18.6 \\ (15.1--22.2)} 
& \shortstack{42.8 \\ (39.1--46.4)} 
& \shortstack{48.3 \\ (44.7--51.9)} 
& \shortstack{48.0 \\ (44.1--52.0)} 
& \shortstack{49.7 \\ (45.9--53.6)} 
& \shortstack{47.9 \\ (43.8--51.8)} 
& \shortstack{\textbf{41.2} \\ (\textbf{37.4-44.8})} 
& <0.01 
& <0.01 \\
\midrule
INbreast
& \shortstack{16.7 \\ (6-29.4)} 
& \shortstack{71.4 \\ (56.1--85.8)} 
& \shortstack{67.4 \\ (52.4--81.7)} 
& \shortstack{64.7 \\ (47.7--81.4)} 
& \shortstack{68.5 \\ (53.5--83.4)} 
& \shortstack{65.1 \\ (50.0--79.0)} 
& \shortstack{\textbf{63.0} \\ (\textbf{44.8-79.3})} 
& <0.01 
& <0.01 \\
\bottomrule
\end{tabular}%
}
\label{tab:Mammo_Seg}
\end{table*}

\begin{table*}
\centering
\caption{Summary of medical ultrasound datasets used in this study. The collection encompasses diverse anatomical regions and clinical tasks.}
\small
\begin{tabular}{l l l cc}
\toprule
\multirow{2}{*}{\textbf{Dataset}} 
& \multirow{2}{*}{\textbf{Task}} 
& \multirow{2}{*}{\textbf{Specific Task}} 
& \multicolumn{2}{c}{\textbf{\#Images}} \\
\cmidrule(lr){4-5}
& & & \textbf{Train} & \textbf{Valid} \\
\midrule
TG3K~\cite{gong2023thyroid}
& Segmentation & Thyroid Nodule Segmentation & 14 & 6 \\

TN3K~\cite{gong2021multi}
& Segmentation & Thyroid Nodule Segmentation & 14 & 6 \\

DDTI~\cite{pedraza2015open}
& Segmentation & Thyroid Segmentation & 80 & 20 \\

MMOTU~\cite{zhao2022multi}
& Segmentation & Liver Segmentation & 80 & 20 \\
\bottomrule
\end{tabular}
\label{tab:UltrasoundDatasets}
\end{table*}

\begin{table*}
\centering
\small 
\caption{Comparison of segmentation performance for ultrasound image datasets. The table presents results for the TG3K, TN3K, DDTI and MMOTU datasets, measured by Dice Similarity Coefficient (DSC) in percentage.}
\resizebox{\linewidth}{!}{%
\begin{tabular}{lccccccc}
\toprule
\textbf{Dataset} 
& \textbf{UNet (Scratch)} 
& \textbf{LVM-Med (R50)} 
& \textbf{LVM-Med (ViT)} 
& \textbf{MedSAM} 
& \textbf{RaSD-2D (Ours)} 
& \textbf{p vs Scratch}
& \textbf{p vs Best} \\
\midrule
TG3K
& \shortstack{93.7 \\ (93.3--94.0)} 
& \shortstack{94.4 \\ (94.2--94.6)} 
& \shortstack{95.3 \\ (95.2--95.4)} 
& \shortstack{95.3 \\ (95.2--95.4)} 
& \shortstack{\textbf{94.6} \\ (\textbf{94.4--94.8})} 
& <0.01 
& <0.01 \\
\midrule
TN3K
& \shortstack{73.0 \\ (71.7--74.3)} 
& \shortstack{74.1 \\ (72.7--75.6)} 
& \shortstack{79.6 \\ (78.5--80.6)} 
& \shortstack{80.1 \\ (79.0--81.1)} 
& \shortstack{\textbf{73.6} \\ (\textbf{72.3--74.9})} 
& 0.06
& <0.01 \\
\midrule
DDTI
& \shortstack{78.2 \\ (76.7--79.7)} 
& \shortstack{80.6 \\ (79.0--82.3)} 
& \shortstack{83.8 \\ (82.3--85.2)} 
& \shortstack{84.4 \\ (83.0--85.9)} 
& \shortstack{\textbf{81.3} \\ (\textbf{79.5--83.0})} 
& <0.01
& <0.01 \\
\midrule
MMOTU
& \shortstack{75.3 \\ (73.9--76.8)} 
& \shortstack{77.4 \\ (76.2--78.7)} 
& \shortstack{82.0 \\ (80.9--83.1)} 
& \shortstack{82.5 \\ (81.4--83.7)} 
& \shortstack{\textbf{80.0} \\ (\textbf{78.7--81.2})} 
& <0.01
& <0.01 \\
\bottomrule
\end{tabular}%
}
\label{tab:Ultrasound_Seg}
\end{table*}

\subsubsection{CT Tasks}
A total of 14 CT datasets were used in this study, including 12 for segmentation tasks and 2 for classification tasks. The detailed information for these datasets is provided in Table~\ref{tab:CTDatasets}. The segmentation performance across 12 datasets is summarized in Table~\ref{tab:CT_seg}, while the classification results are presented in Table~\ref{tab:CT_MR_Classification}.

\textbf{TotalSegmentator}\cite{wasserthal2023totalsegmentator} The TotalSegmentator dataset is a large-scale CT imaging dataset designed for automated segmentation of major anatomical structures. We use its version 2 in our experiment. It contains 1,228 clinical CT scans with high-quality manual annotations for 104 structures, including organs, bones, muscles, and vessels.

\textbf{BHSD}\cite{wu2023bhsd} The Brain Hemorrhage Segmentation Dataset (BHSD) provides a 3D multiclass ICH dataset containing 192 volumes with pixel-level annotations and 2200 volumes with slice-level annotations across five categories of intracranial hemorrhage.

\textbf{BTCV}\cite{landman2015miccai} The Beyond the Cranial Vault (BTCV) abdomen challenge dataset is provided by the Vanderbilt University Medical Center and consists of 30 abdominal CT scans with annotations for 13 different organs, which are spleen, right kidney, left kidney, gallbladder, esophagus, liver, stomach, aorta, inferior vena cava, portal and splenic veins, pancreas, left and right adrenal glands. 

\textbf{IRCADb}\cite{soler20103d} The IRCADb dataset is a publicly available medical imaging database developed by the French Institute for Research Against Digestive Cancer (IRCAD). It consists of 20 contrast-enhanced abdominal CT scans, including 10 female and 10 male subjects
, with hepatic tumors present in 75\% of the cases.

\textbf{CHAOS}\cite{kavur2021chaos} The Combined (CT-MR) Healthy Abdominal Organ Segmentation challenge dataset. For our CT segmentation experiments, we utilize the abdominal CT subset containing 40 scans.

\textbf{MSD}\cite{antonelli2022medical} Medical Segmentation Decathlon (MSD) comprised different target regions, modalities and challenging characteristics and was separated into ten tasks. In our experiment, MSD Task 07 and MSD Task 09 was used for the segmentation of Pancreas Tumor and Spleen.

\textbf{Covid-19-20}\cite{roth2022rapid} COVID-19 Lung CT Lesion Segmentation Challenge - 2020 (Covid-19-20) dataset contains chest CT scans of COVID-19 patients with expert annotations of lung lesions.

\textbf{Parse22}\cite{luo2023efficient} Pulmonary Artery Segmentation Challenge 2022 (Parse22) contains 200 3D volumes with refined pulmonary artery labeling from 10 clinicians.

\textbf{SegThor}\cite{lambert2020segthor} The Segmentation of THoracic Organs at Risk (SegThor) includes 60 3D CT scans and the the segmentation of organs at risk have been contoured manually by an experienced radiotherapist, which includes the heart, the trachea, the aorta and the esophagus.

\textbf{MM-WHS}\cite{zhuang2018multivariate} The Multi-Modality Whole Heart Segmentation (MM-WHS) dataset was developed for the MICCAI 2017 challenge and features 40 multi-modal cardiac images, consisting of 20 cardiac CT volumes and 20 cardiac MRI volumes. The dataset includes manual segmentation for seven major cardiac substructures, the left and right ventricular cavities, the left and right atrial cavities, the left ventricular myocardium, the ascending aorta, and the pulmonary artery.

\textbf{VerSe20}\cite{sekuboyina2021verse}  Large Scale Vertebrae Segmentation Challenge (VerSe) is a large scale, multi-detector, multi-site, CT spine dataset consisting of 374 scans from 355 patients.

\textbf{CC-CCII}\cite{zhang2020clinically} The China Consortium of Chest CT Image Investigation (CC-CCII) dataset comprises 4178 CT chest scans. It categorizes the images into three main groups: novel coronavirus pneumonia (NCP) caused by SARS-CoV-2 infection, common pneumonia (CP), and normal controls(Normal). 

\textbf{LUNA16}\cite{setio2017validation} The LUNA16 (LUng Nodule Analysis 2016) dataset is derived from the LIDC-IDRI database and contains 888 annotated chest CT scans. We densely extract 3D patches from each CT volume using a sliding window approach, generating approximately 621 patches per CT on average. Given that positive samples constitute only about 0.2\% of the total data, each positive training sample is augmented to generate 10 variants.

\subsubsection{MR Tasks} 

A total of 6 MR datasets were used in this study. The detailed information for these datasets is provided in Table~\ref{tab:MRDatasets}. Five datasets (BraTS21, AMOS-MR, ATLAS-MR, WHS-MR, and ACDC) are used for segmentation tasks, with their results summarized in Table~\ref{tab:MR_Seg}. An additional dataset (MRNet) is employed for classification tasks, with results summarized in Table~\ref{tab:CT_MR_Classification}.

\textbf{BraTS21}\cite{baid2021rsna} BraTs21 dataset is an MRI dataset, containing 1251 scans with classes: whole tumor (WT), tumor core (TC), and enhancing tumor (ET).

\textbf{AMOS}\cite{ji2022amos} Multi-Modality Abdominal MultiOrgan Segmentation Challenge (AMOS) provides 500 CT and 100 MRI scans collected from multicenter, multi-vendor, multi-modality, multi-phase, multi-disease patients, each with voxel-level annotations of 15 abdominal organs. In our experiment, only the MRI images are used.

\textbf{ATLAS}\cite{quinton2023tumour} The Anatomical Tracings of Lesions after Stroke Dataset (ATLAS) consisting a total of 955 T1-weighted MRIs with manually segmented diverse lesions and metadata.

\textbf{ACDC}\cite{bernard2018deep} The Automatic Cardiac Diagnosis Challenge (ACDC) dataset is a cardiac MRI dataset featuring 150 subjects with expert annotations for the left ventricle, myocardium, and right ventricle. 

\textbf{MRNet} \cite{bien2018deep} The MRNet dataset comprises 1,370 exams from Stanford University Medical Center. It contains labels for abnormal studies, anterior cruciate ligament (ACL) tears, and meniscal tears.

\subsubsection{X-ray Tasks}

A total of 16 X-ray datasets were used in this study. Six datasets (Shenzhen set, MC set, RSNA-PBA, BFMR) are used for chest X-ray tasks, with segmentation and classification results summarized in Table~\ref{tab:Xray_Seg} and Table~\ref{tab:Xray_Cls}, respectively. Additionally, 10 mammography datasets (KAU-BCMD, BMCD, etc.) are employed for classification, detection, and segmentation tasks, with their results detailed in Tables~\ref{tab:Mammo_Cls},\ref{tab:Mammo_Det}, and \ref{tab:Mammo_Seg}. The detailed information for all datasets is provided in Tables~\ref{tab:XrayDatasets} and \ref{tab:MammoDatasets}.

\textbf{RSNA-PBA} \cite{halabi2019rsna} 
The RSNA Pediatric Bone Age Challenge (RSNA-PBA) is a public dataset for pediatric bone age assessment from hand radiographs. It contains 14,236 2D hand X-ray images with corresponding bone age annotations.

\textbf{Shenzhen set} \cite{jaeger2014two}
The Shenzhen Chest X-ray Set (Shenzhen set) is a public chest X-ray dataset for pulmonary tuberculosis screening, containing 662 X-rays. We utilized its image-level labels for tuberculosis classification and the provided lung segmentation masks for model training and evaluation.

\textbf{MC set} \cite{jaeger2014two}
The Montgomery County Chest X-ray Set (MC set)  is a public chest X-ray dataset for pulmonary tuberculosis screening, consisting of 138 X-rays. We utilized its image-level labels for tuberculosis classification and the provided lung segmentation masks for model training and evaluation.

\textbf{BFMR} \cite{fracturexray}
The Bone Fracture Multi-Region (BFMR) X-ray Dataset comprises 10,580 X-ray images captured from various anatomical regions, including the upper and lower limbs, spine, and hip. 

\textbf{KAU-BCMD} \cite{alsolami2021king}
The King Abdulaziz University Breast Cancer Mammogram Dataset (KAU-BCMD) provides a comprehensive collection of 1,416 cases (5,662 mammograms) with annotations including BI-RADS assessments, tumor masks, and glandular tissue percentage. In this study, we specifically utilized the percentage of glandular tissue to address the breast composition classification task.

\textbf{BMCD} \cite{loizidou2021breast}
Breast Micro-Calcifications Dataset (BMCD) comprises 100 pairs of sequential screening mammograms (400 images in total) that include normal, benign, and suspicious cases. It provides radiologist-annotated labels for each micro-calcification, BI-RADS assessments, and breast density information. In this study, we utilized the breast density for composition classification and the BI-RADS categories for malignancy likelihood assessment.

\textbf{LAMIS-DMDB} \cite{imane2024lamis}
The Laboratory of Mathematics, Informatics and System
Digital Mammogram Data Base (LAMIS-DMDB) is a large-scale public dataset containing 2,216 mammogram exams. We utilized its BI-RADS labels for classification.

\textbf{CDD-CESM} \cite{khaled2022categorized}
The Categorized Digital Database for Low-energy and Subtracted Contrast-Enhanced Spectral Mammography Images (CDD-CESM) provides paired low-energy and contrast-enhanced spectral mammograms. 

\textbf{MM} \cite{aqdar2024mammogram}
The Mammographic Mass (MM) Dataset is a collection of 745 mammogram images with binary labels indicating breast cancer presence.

\textbf{INbreast} \cite{moreira2012inbreast}
The INbreast dataset is a public mammography database containing 410 images with expert annotations for masses and calcifications. In this study, we utilized this dataset for breast lesion detection and segmentation experiments. For the detection task, bounding boxes were generated by calculating the minimum enclosing rectangles of the finely annotated lesion contours.

\textbf{VinDr-Mammo} \cite{nguyen2023vindr}
The VinDr-Mammo Dataset is a large-scale public dataset for detecting and categorizing breast lesions in mammograms, providing both image-level and bounding box annotations for a variety of lesions.

\textbf{CBIS-DDSM} \cite{lee2017curated}
The Curated Breast Imaging Subset of the Digital Database for Screening Mammography (CBIS-DDSM) contains 1644 curated cases (753 calcification and 891 mass cases). It provides pixel-level segmentation masks of lesions annotated by radiologists.

\subsubsection{Ultrasound Task}
A total 4 ultrasound datasets were used in this study. All datasets are used for segmentation tasks, with their results summarized in Table~\ref{tab:Ultrasound_Seg}. The detailed information for all datasets is provided in Table~\ref{tab:UltrasoundDatasets}.

\textbf{TG3K} \cite{gong2023thyroid} The TG3K dataset is a 2D ultrasound dataset constructed from 16 thyroid ultrasound videos by extracting image frames and applying a region coverage criterion. It consists of 3,583 high-quality images with annotations and is designed for thyroid nodule segmentation.

\textbf{TN3K} \cite{gong2021multi} The Thyroid Nodule Region Segmentation Dataset (TN3K) is a public dataset for thyroid nodule segmentation in ultrasound images. It consists of 3,493 2D ultrasound images collected from 2,421 patients. All images contain at least one thyroid nodule and are annotated for segmentation.

\textbf{DDTI} \cite{pedraza2015open} The Digital Database of Thyroid Ultrasound Images (DDTI) is a public dataset for thyroid nodule segmentation in ultrasound images. It contains 637 2D ultrasound images acquired from a single device, with pixel-level annotations provided by clinical experts. The dataset includes diverse thyroid conditions, such as thyroiditis, goiters, nodules, and cancer.

\textbf{MMOTU} \cite{zhao2022multi} The Multi-Modality Ovarian Tumor Ultrasound (MMOTU) dataset is a public ultrasound dataset for ovarian tumor segmentation. It consists of two subsets corresponding to different ultrasound modalities, including 2D ultrasound images (OTU\_2D) and contrast-enhanced ultrasound images. In this work, we only use the OTU\_2D subset, which contains 1,469 2D ultrasound images with pixel-wise annotations for ovarian tumor segmentation.

\begin{table*}[ht]
\centering
\caption{Summary of medical fundus datasets used in this study. The collection encompasses diverse tasks and prediction targets.}
\small
\begin{tabular}{l l l cc}
\toprule
\multirow{2}{*}{\textbf{Dataset}} 
& \multirow{2}{*}{\textbf{Task}} 
& \multirow{2}{*}{\textbf{Specific Task}} 
& \multicolumn{2}{c}{\textbf{\#Images}} \\
\cmidrule(lr){4-5}
& & & \textbf{Train} & \textbf{Valid} \\
\midrule
STARE~\cite{hoover2000locating}
& Segmentation & Vessel Segmentation & 14 & 6 \\

DRIVE~\cite{staal2004ridge}
& Segmentation & Vessel Segmentation & 14 & 6 \\

HVDROPDB-BV~\cite{agrawal2024hvdropdb}
& Segmentation & Vessel Segmentation & 80 & 20 \\

HVDROPDB-OD~\cite{agrawal2024hvdropdb}
& Segmentation & Optic Disc Segmentation & 80 & 20 \\

HVDROPDB-RIDGE~\cite{agrawal2024hvdropdb}
& Segmentation & Ridge Segmentation & 80 & 30 \\

\midrule

IDRiD~\cite{porwal2018indian}
& Classification & Macular Edema Risk Classification & 295 & 160 \\

DRAC~\cite{qian2024drac}
& Classification & DR Grading & 499 & 112 \\
\bottomrule
\end{tabular}
\label{tab:FundusDatasets}
\end{table*}

\begin{table*}
\centering
\small 
\caption{Comparison of segmentation performance for fundus image datasets. The table presents results for the STARE, DRIVE, and HVDROPDB datasets, measured by Dice Similarity Coefficient (DSC) in percentage.}
\resizebox{\linewidth}{!}{%
\begin{tabular}{lccccccc}
\toprule
\textbf{Dataset} 
& \textbf{UNet (Scratch)} 
& \textbf{LVM-Med (R50)} 
& \textbf{LVM-Med (ViT)} 
& \textbf{MedSAM} 
& \textbf{RaSD-2D (Ours)} 
& \textbf{p vs Scratch}
& \textbf{p vs Best} \\
\midrule
STARE
& \shortstack{79.3 \\ (75.1--83.5)} 
& \shortstack{80.1 \\ (75.2--85.0)} 
& \shortstack{86.1 \\ (83.3--89.0)} 
& \shortstack{85.0 \\ (81.6--88.5)} 
& \shortstack{\textbf{80.4} \\ (\textbf{77.7-83.1})} 
&  0.26
&  <0.01 \\
\midrule
DRIVE
& \shortstack{82.3 \\ (79.0--85.6)} 
& \shortstack{86.0 \\ (83.2--88.8)} 
& \shortstack{87.7 \\ (87.0--88.6)} 
& \shortstack{88.0 \\ (87.2--88.8)} 
& \shortstack{\textbf{86.7} \\ (\textbf{85.1--88.3})} 
& <0.01
&  0.03\\
\midrule
HVDROPDB-BV
& \shortstack{64.7 \\ (61.1--68.3)} 
& \shortstack{67.5 \\ (64.5--70.4)} 
& \shortstack{76.4 \\ (75.0--77.8)} 
& \shortstack{77.4 \\ (76.2--78.7)} 
& \shortstack{\textbf{68.4} \\ (\textbf{65.8--70.9})} 
& <0.01
& <0.01 \\
\midrule
HVDROPDB-OD
& \shortstack{71.2 \\ (62.3--80.1)} 
& \shortstack{77.7 \\ (70.4--84.9)} 
& \shortstack{85.2 \\ (77.1--93.4)} 
& \shortstack{70.9 \\ (63.4--78.4)} 
& \shortstack{\textbf{65.2} \\ (\textbf{68.0--82.4})} 
& 0.38
& 0.02 \\
\midrule
HVDROPDB-RIDGE
& \shortstack{52.1 \\ (50.1--54.2)} 
& \shortstack{66.3 \\ (58.9--73.6)} 
& \shortstack{66.9 \\ (63.2--70.6)} 
& \shortstack{65.1 \\ (61.4--68.8)} 
& \shortstack{\textbf{63.1} \\ (\textbf{57.0--69.1})} 
& <0.01
& 0.30 \\
\bottomrule
\end{tabular}%
}
\label{tab:Fundus_Seg}
\end{table*}

\begin{table*}
\centering
\small 
\caption{Comparison of classification performance for fundus image datasets. The table presents results for the HVDROPDB, IDRiD, and DRAC datasets, measured by AUC in percentage.}
\resizebox{\linewidth}{!}{%
\begin{tabular}{lccccccc}
\toprule
\textbf{Dataset} 
& \textbf{UNet (Scratch)} 
& \textbf{LVM-Med (R50)} 
& \textbf{LVM-Med (ViT)} 
& \textbf{MedSAM} 
& \textbf{RaSD-2D (Ours)} 
& \textbf{p vs Scratch}
& \textbf{p vs Best} \\
\midrule
DRAC
& \shortstack{67.6 \\ (62.4--72.8)} 
& \shortstack{77.2 \\ (60.8--93.7)} 
& \shortstack{81.5 \\ (76.7--86.2)} 
& \shortstack{85.3 \\ (79.4--91.2)} 
& \shortstack{\textbf{71.5} \\ (\textbf{58.3--84.7})} 
& <0.01
& <0.01\\
\midrule
IDRiD
& \shortstack{70.9 \\ (65.5--76.4)} 
& \shortstack{88.5 \\ (83.7--93.2)} 
& \shortstack{82.4 \\ (76.7--86.2)} 
& \shortstack{83.2 \\ (79.4--91.2)} 
& \shortstack{\textbf{74.1} \\ (\textbf{58.3--84.7})} 
& <0.01
& <0.01\\
\bottomrule
\end{tabular}%
}
\label{tab:Fundus_Cls}
\end{table*}

\begin{table*}
\centering
\caption{Summary of medical pathology datasets used in this study. The collection encompasses diverse anatomical regions and clinical tasks.}
\small
\begin{tabular}{l l l c cc}
\toprule
\multirow{2}{*}{\textbf{Dataset}}  
& \multirow{2}{*}{\textbf{Task}}
& \multirow{2}{*}{\textbf{Anatomical Region}}
& \multirow{2}{*}{\textbf{\#Classes}} 
& \multicolumn{2}{c}{\textbf{\#Images}} \\
\cmidrule(lr){5-6}
& & & & \textbf{Train} & \textbf{Valid} \\
\midrule
GlaS~\cite{sirinukunwattana2017gland}
& Segmentation & Colon & 1 & 85 & 80 \\

CoCaHis~\cite{sitnik2021dataset}
& Segmentation & Liver & 1 & 58 & 24 \\

TNBC~\cite{naylor2018segmentation}
& Segmentation & Breast & 11 & 54 & 14 \\

MoNuSAC~\cite{verma2021monusac2020}
& Segmentation & Whole Body & 4 & 242 & 117 \\

Janowczyk~\cite{janowczyk2016deep}
& Segmentation & Breast & 1 & 451 & 113 \\

CRAG~\cite{graham2019mild}
& Segmentation & Colon & 1 & 1557 & 362 \\

MIDOG~\cite{aubreville2024domain}
& Segmentation & Trunk & 2 & 1185 & 297 \\

NuCLS~\cite{amgad2022nucls}
& Segmentation & Breast & 12 & 1395 & 349 \\

WSSS4LUAD~\cite{han2022wsss4luad}
& Segmentation & Lung & 3 & 2290 & 573 \\
\bottomrule
\end{tabular}
\label{tab:PathDatasets}
\end{table*}

\begin{table*}
\centering
\small 
\caption{Comparison of segmentation performance for histopathology image datasets. The table presents results for nine public datasets, including GlaS, CoCaHis, and TNBC, measured by Dice Similarity Coefficient (DSC) in percentage.}
\resizebox{\linewidth}{!}{%
\begin{tabular}{lcccccccc}
\toprule
\textbf{Dataset} 
& \textbf{UNet (Scratch)} 
& \textbf{LVM-Med (R50)} 
& \textbf{LVM-Med (ViT)} 
& \textbf{MedSAM} 
& \textbf{PathSegmentor} 
& \textbf{RaSD-2D (Ours)} 
& \textbf{p vs Scratch}
& \textbf{p vs Best} \\
\midrule
GlaS
& \shortstack{90.9 \\ (89.2--92.6)} 
& \shortstack{90.8 \\ (89.3--92.3)} 
& \shortstack{90.6 \\ (88.7--92.5)} 
& \shortstack{88.4 \\ (84.1--90.7)} 
& \shortstack{93.7 \\ (92.5--94.9)} 
& \shortstack{\textbf{91.5} \\ (\textbf{90.1--92.9})} 
& 0.06
& <0.01 \\
\midrule
CoCaHis
& \shortstack{77.8 \\ (75.5--80.1)} 
& \shortstack{85.4 \\ (83.9--86.9)} 
& \shortstack{84.4 \\ (81.9--86.9)} 
& \shortstack{83.7 \\ (81.0--86.4)} 
& \shortstack{82.7 \\ (80.7--84.7)} 
& \shortstack{\textbf{88.0} \\ (\textbf{87.0--89.0})} 
& <0.01
& <0.01 \\
\midrule
TNBC
& \shortstack{37.9 \\ (31.2--44.6)} 
& \shortstack{37.1 \\ (29.6--44.6)} 
& \shortstack{40.1 \\ (32.7--47.5)} 
& \shortstack{33.5 \\ (26.1--40.9)} 
& \shortstack{43.0 \\ (35.5--50.5)} 
& \shortstack{\textbf{41.6} \\ (\textbf{34.1--49.1})} 
& 0.16
& 0.54 \\
\midrule
MoNuSAC
& \shortstack{59.0 \\ (53.2--64.8)} 
& \shortstack{59.0 \\ (53.5--64.5)} 
& \shortstack{61.6 \\ (55.9--67.3)} 
& \shortstack{59.6 \\ (53.4--65.8)} 
& \shortstack{65.7 \\ (59.5--71.9)} 
& \shortstack{\textbf{61.4} \\ (\textbf{55.8--67.0})} 
& 0.02
& 0.85\\
\midrule
Janowczyk
& \shortstack{41.2 \\ (36.9--45.5)} 
& \shortstack{39.0 \\ (34.9--43.1)} 
& \shortstack{40.1 \\ (36.2--44.0)} 
& \shortstack{38.7 \\ (34.8--42.6)} 
& \shortstack{34.0 \\ (30.2--37.8)} 
& \shortstack{\textbf{42.1} \\ (\textbf{37.8--46.4})} 
& 0.27
& <0.01 \\
\midrule
CRAG
& \shortstack{89.2 \\ (86.8--91.6)} 
& \shortstack{89.2 \\ (86.8--91.6)} 
& \shortstack{91.9 \\ (90.2--93.6)} 
& \shortstack{91.5 \\ (89.5--93.5)} 
& \shortstack{93.3 \\ (91.8--94.8)} 
& \shortstack{\textbf{90.3} \\ (\textbf{88.2--92.4})} 
& <0.01
& <0.01 \\
\midrule
MIDOG
& \shortstack{26.6 \\ (19.3--33.9)} 
& \shortstack{24.2 \\ (17.0--31.4)} 
& \shortstack{26.8 \\ (19.4--34.2)} 
& \shortstack{29.2 \\ (22.2--36.2)} 
& \shortstack{38.8 \\ (30.5--47.1)} 
& \shortstack{\textbf{27.3} \\ (\textbf{20.1--34.5})} 
& 0.52
& 0.21 \\
\midrule
NuCLS
& \shortstack{25.6 \\ (20.0--31.2)} 
& \shortstack{25.6 \\ (20.3--30.9)} 
& \shortstack{27.5 \\ (21.8--33.2)} 
& \shortstack{25.8 \\ (22.2--31.4)} 
& \shortstack{37.1 \\ (30.5--47.1)} 
& \shortstack{\textbf{27.4} \\ (\textbf{20.1--34.5})} 
& <0.01
& 0.83 \\
\midrule
WSSS4LUAD
& \shortstack{82.5 \\ (78.4--86.6)} 
& \shortstack{82.3 \\ (78.3--86.3)} 
& \shortstack{84.0 \\ (79.9--88.1)} 
& \shortstack{83.4 \\ (79.2--87.6)} 
& \shortstack{83.3 \\ (79.3--87.3)} 
& \shortstack{\textbf{83.0} \\ (\textbf{79.0--87.0})} 
& <0.01
& <0.01 \\
\bottomrule
\end{tabular}%
}
\label{tab:Path_seg}
\end{table*}

\subsubsection{Fundus Task}

A total of 8 fundus datasets were used in this study. Four datasets (STARE, DRIVE, and the multi-task HVDROPDB dataset with its BV, OD, and RIDGE segmentation tasks) are used for segmentation, with results summarized in Table~\ref{tab:Fundus_Seg}. Three datasets (HVDROPDB with its classification task, IDRiD, and DRAC) are employed for classification tasks, with results summarized in Table~\ref{tab:Fundus_Cls}. The detailed information for all datasets is provided in Table~\ref{tab:FundusDatasets}.

\textbf{STARE} \cite{hoover2000locating}
The STructured Analysis of the Retina (STARE) dataset is a foundational resource for blood vessel segmentation. It consists of 20 color fundus images, with approximately half containing various pathologies.

\textbf{DRIVE} \cite{staal2004ridge}
The Digital Retinal Images for Vessel Extraction (DRIVE) dataset contains 20 color fundus images with pixel-level segmentation mask. 

\textbf{HVDROPDB} \cite{agrawal2024hvdropdb}
The HVDROPDB dataset from H.V. Desai Hospital is a comprehensive resource for Retinopathy of Prematurity (ROP) analysis. It includes two main components: three fine-grained segmentation subsets (HVDROPDB-OD, -BV, -RIDGE), each with 50 images and masks from both RetCam and Neo systems for the optic disc, blood vessels, and ridge, respectively. In our experiments, images from both RetCam and Neo across all tasks were combined for model training.

\textbf{IDRiD} \cite{porwal2018indian}
The Indian Diabetic Retinopathy Image Dataset (IDRiD) comprises 516 high-resolution fundus images from real-world clinical practice in India. Each image is graded for disease severity according to international standards. For our study, we specifically utilized the expert-assigned severity levels for Diabetic Macular Edema for classification.

\textbf{DRAC} \cite{qian2024drac}
The DRAC (Diabetic Retinopathy Analysis Challenge) dataset was employed for diabetic retinopathy grading, containing 611 fundus images with labels. These images are classified into three clinical grades: No DR, Non-Proliferative DR, and Proliferative DR.

\subsubsection{Pathology Task} 

A total of 9 pathology datasets were used in this study. All datasets are used for segmentation tasks, with their results summarized in Table~\ref{tab:Path_seg}. The detailed information for all datasets is provided in Table~\ref{tab:PathDatasets}.

\textbf{GlaS} \cite{sirinukunwattana2017gland}
Gland Segmentation in Colon Histology Images Challenge Contest (GlaS) dataset is a public dataset for gland instance segmentation in H\&E-stained colon histology images, introduced as part of the MICCAI 2015 challenge contest. It contains 165 images with expert-annotated instance segmentation masks for glands.

\textbf{CoCaHis} \cite{sitnik2021dataset}
The Colon Cancer Histopathological (CoCaHis) dataset is designed for intraoperative diagnosis of metastatic colon cancer in liver tissue. It contains 82 histopathology images from 19 patients, with pixel-wise annotations distinguishing cancerous regions.

\textbf{TNBC} \cite{naylor2018segmentation}
The Triple Negative Breast Cancer (TNBC) dataset consists of annotated H\&E stained histology images at 40× magnification. The dataset provides instance-level masks for over 4,000 cell nuclei.

\textbf{MoNuSAC} \cite{verma2021monusac2020}
The Multi-organ Nuclei Segmentation and Classification (MoNuSAC) dataset comprises H\&E-stained histology images from 37 hospitals, involving 71 patients. The dataset features over 46,000 manually annotated nuclei, which are categorized into four classes (epithelial cells, lymphocytes, macrophages, and neutrophils).

\textbf{Janowczyk} \cite{janowczyk2016deep}
The Janowczyk dataset contains 141 image patches at 40x magnification, with meticulously annotated instance-level masks for approximately 12,000 cell nuclei.

\textbf{CRAG} \cite{graham2019mild}
The Colorectal Adenocarcinoma Gland (CRAG) comprises H\&E-stained whole slide images obtained from the University Hospitals Coventry and Warwickshire (UHCW) NHS Trust, is designed for gland instance segmentation. It provides high-quality, instance-level annotations of glands for this purpose.

\textbf{MIDOG} \cite{aubreville2024domain}
The 2022 challenge on MItosis Domain Generalization (MIDOG 2022) dataset is designed to address the challenge of domain generalization in mitotic figure detection across diverse histopathology images. It provides annotated images from six different domains, encompassing various tumor types, laboratories, and species.

\textbf{NuCLS} \cite{amgad2022nucls}
The nucleus classification, localization, and segmentation (NuCLS) dataset is a large-scale, crowdsourced benchmark for nucleus-related tasks in breast cancer. It contains over 220,000 annotated nuclei from breast cancer images sourced from The Cancer Genome Atlas (TCGA).

\textbf{WSSS4LUAD} \cite{han2022wsss4luad}
The WSSS4LUAD dataset was introduced through the Weakly-supervised Tissue Semantic Segmentation for Lung Adenocarcinoma grand challenge. The dataset provides H\&E-stained whole slide images (WSIs) with only image-level labels indicating the presence of epithelium, stroma, and normal tissue.

\begin{figure*}
    \centering
    \includegraphics[width=0.9\linewidth]{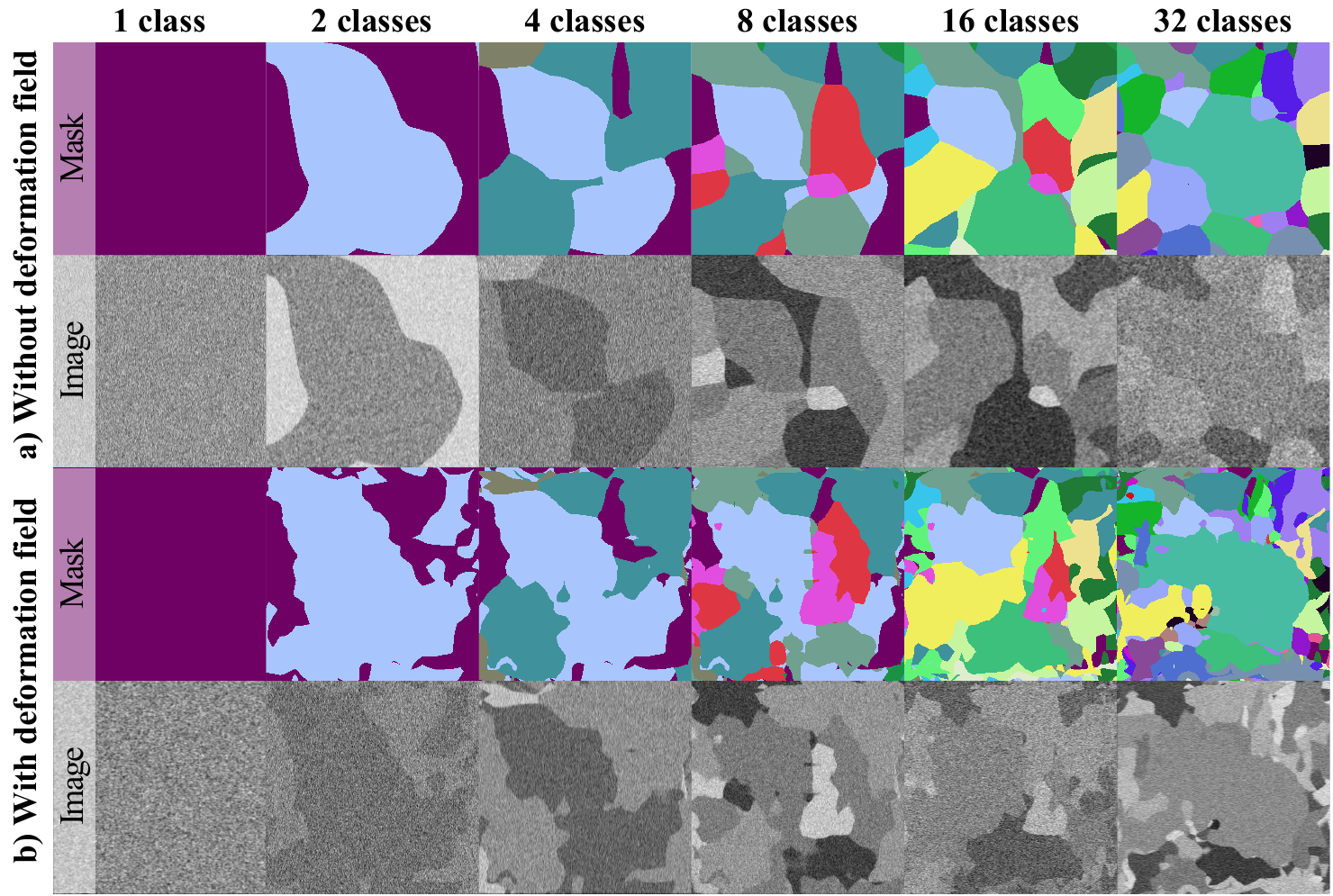}
    \caption{The visualization of our synthetic image-mask pairs in our data engine with different class amounts. a) Without the deformation field, the structures tend to be whole blocks and lack thin geometry. b) When adding the deformation field, the structures tend to be diversified, improving the geometry diversity.}
    \label{fig:synthetic_data}
\end{figure*}

\subsection{Visualization}

\subsubsection{Visualization of synthesis} As shown in Fig.~\ref{fig:synthetic_data}, we visualize the generated synthetic image and mask pairs with different class amounts (in 2D slice). There are two observations: 1) In the case of a fixed image size, the more categories there are, the more trivial the region in the image will be, and the richer the information of small structures will be. Through our image synthesis engine, these masks are filled with Gaussian noise with different means and standard deviations (stds), and a synthetic image similar to the medical image style is obtained. 2) The deformation field in our data engine improves geometry diversity, enabling the pre-trained model to adapt to more diverse real anatomical structures. Without the deformation field, the generated structures tend to be whole blocks limited by thin details. When adding the deformation field, the structures are much more diverse.

\subsubsection{Visualization of features} 
\begin{figure*}
    \centering
    \includegraphics[width=\linewidth]{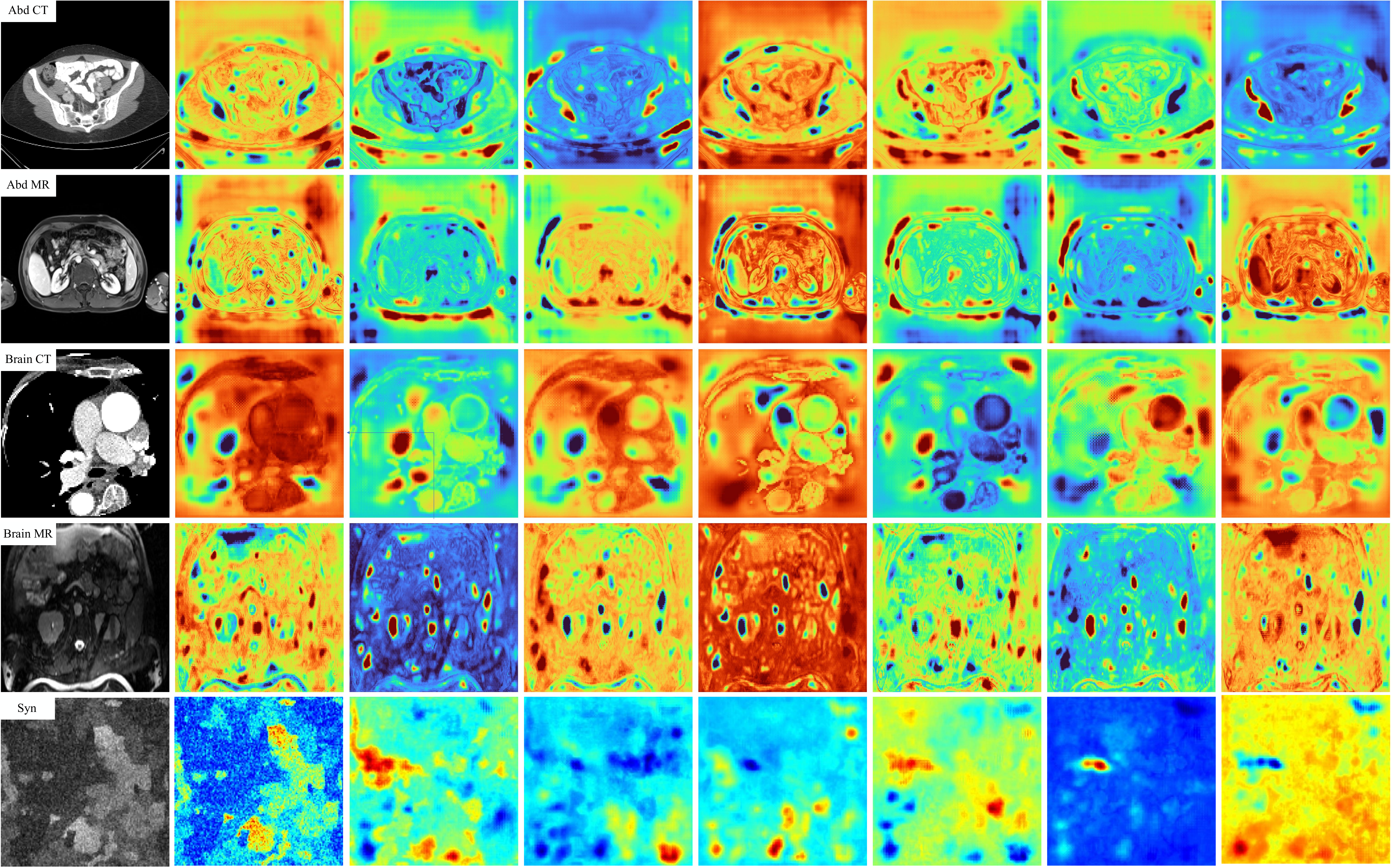}
    \caption{The visualization of the feature maps extracted by our RaSD-3D pre-trained model.}
    \label{fig:features_3d}
\end{figure*}

\begin{figure*}
    \centering
    \includegraphics[width=\linewidth]{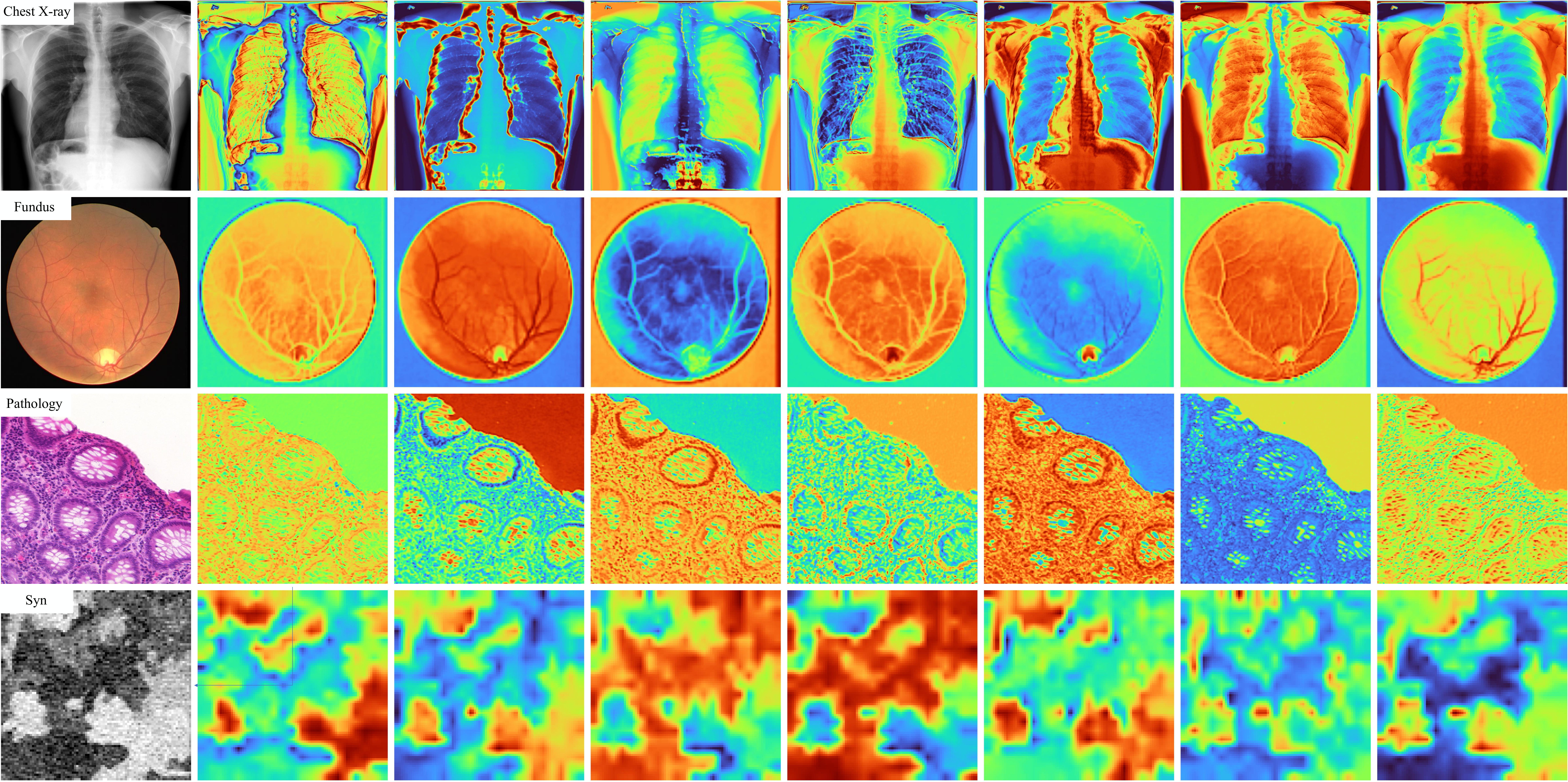}
    \caption{The visualization of the feature maps extracted by our RaSD-2D pre-trained model.}
    \label{fig:features_2d}
\end{figure*}

We visualize the extracted feature maps from our pre-trained RaSD to demonstrate its performance on diverse real-world data. As shown in Fig.\ref{fig:features_2d} and \ref{fig:features_3d}, despite being trained exclusively on synthetic images, our model’s feature representations exhibit strong discriminative power when applied to real medical images. In both abdominal CT and chest MR scans, the feature maps clearly delineate distinct anatomical regions, indicating that the model has effectively learned the semantic structures inherent in medical imagery. This robust discrimination suggests that the synthetic training process has endowed the model with a fundamental understanding of the underlying tissue and organ distributions, even under varying imaging modalities and conditions. As a result, RaSD provides a reliable foundation for a range of downstream tasks such as segmentation and classification, where accurate identification of region-specific features is critical for clinical decision-making.

\subsubsection{Segmentation Results} 
\begin{figure*}
    \centering
    \includegraphics[width=\linewidth]{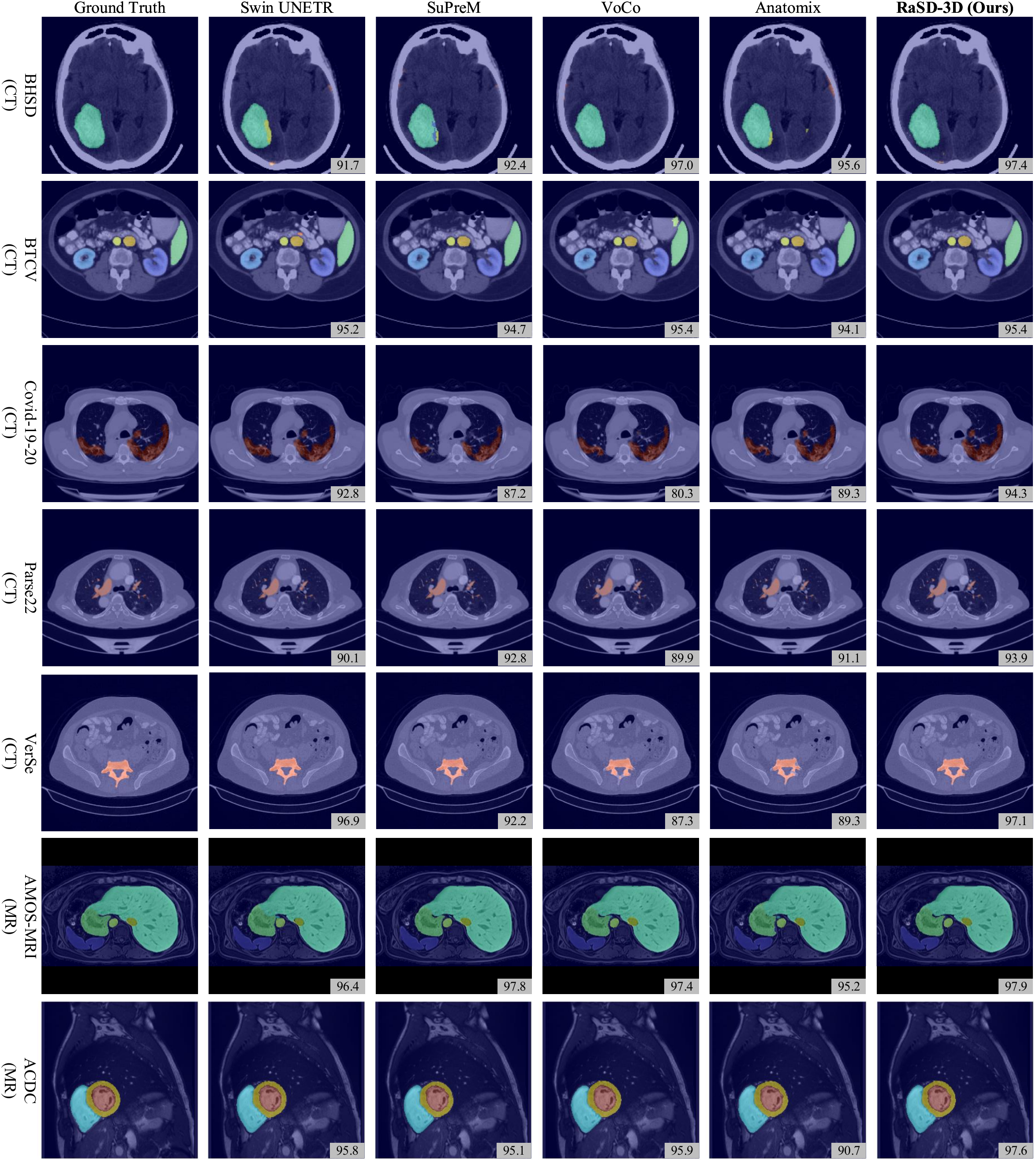}
    \caption{The visualization of the 3D segmentation results.}
    \label{fig:3Dsegmentation}
\end{figure*}

\begin{figure*}
    \centering
    \includegraphics[width=\linewidth]{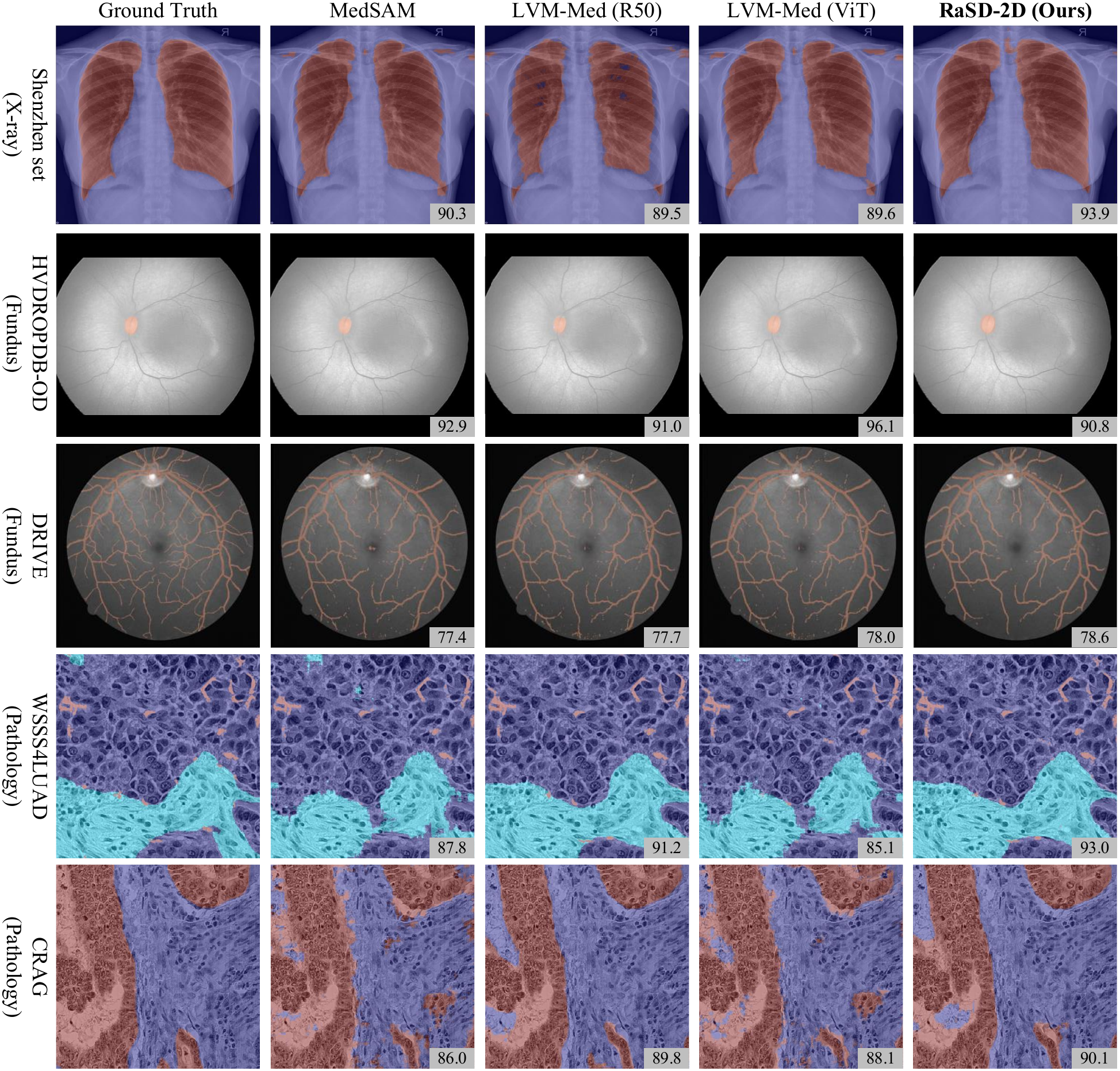}
    \caption{The visualization of the 2D segmentation results.}
    \label{fig:2Dsegmentation}
\end{figure*}

To demonstrate the model’s generalizability across heterogeneous imaging domains, we visualize segmentation outcomes from a diverse set of modalities. As shown in Fig.~\ref{fig:3Dsegmentation}, we include representative results from 3D CT and MR volumes, while Fig.~\ref{fig:2Dsegmentation} presents examples from 2D X-ray, fundus, and pathology images.

\end{document}